\documentclass[traditabstract]{aa}
\usepackage{graphics}
\usepackage{graphicx}
\usepackage{multicol}
\usepackage[varg]{txfonts}

\usepackage{natbib,twoopt}
\usepackage{amsmath} 
\usepackage[breaklinks=true]{hyperref} 
\bibpunct{(}{)}{;}{a}{}{,} 
\makeatletter
\newcommandtwoopt{\citeads}[3][][]{\href{http://adsabs.harvard.edu/abs/#3}%
{\def\hyper@linkstart##1##2{}%
\let\hyper@linkend\@empty\citealp[#1][#2]{#3}}}
\newcommandtwoopt{\citepads}[3][][]{\href{http://adsabs.harvard.edu/abs/#3}%
{\def\hyper@linkstart##1##2{}%
\let\hyper@linkend\@empty\citep[#1][#2]{#3}}}
\newcommandtwoopt{\citetads}[3][][]{\href{http://adsabs.harvard.edu/abs/#3}%
{\def\hyper@linkstart##1##2{}%
\let\hyper@linkend\@empty\citet[#1][#2]{#3}}}
\newcommandtwoopt{\citeyearads}[3][][]%
{\href{http://adsabs.harvard.edu/abs/#3}
{\def\hyper@linkstart##1##2{}%
\let\hyper@linkend\@empty\citeyear[#1][#2]{#3}}}
\makeatother

\usepackage{color}
\definecolor{mygreen}{RGB}{0,128,0}

\hypersetup{colorlinks=true,linkcolor=blue,citecolor=blue,urlcolor=blue}

\begin{document}

\title{The dust disk and companion of the nearby AGB star L$_2$\,Puppis\thanks{Based on observations made with ESO telescopes at Paranal Observatory, under ESO Science Verification program 60.A-9367(A).}}
\subtitle{SPHERE/ZIMPOL polarimetric imaging at visible wavelengths}
\titlerunning{SPHERE/ZIMPOL observations of L$_2$\,Puppis}
\authorrunning{P. Kervella et al.}
%
\author{
P.~Kervella\inst{1,2}
\and
M. Montarg\`es\inst{2,3}
\and
E.~Lagadec\inst{4}
\and
S.~T.~Ridgway\inst{5}
\and
X.~Haubois\inst{6}
\and
J.~H.~Girard\inst{6}
\and
K.~Ohnaka\inst{7}
\and
G.~Perrin\inst{2}
\and
A.~Gallenne\inst{8}
}
\institute{
Unidad Mixta Internacional Franco-Chilena de Astronom\'{i}a (UMI 3386), CNRS/INSU, France
\& Departamento de Astronom\'{i}a, Universidad de Chile, Camino El Observatorio 1515, Las Condes, Santiago, Chile,  \email{pkervell@das.uchile.cl}.
\and
LESIA (UMR 8109), Observatoire de Paris, CNRS, UPMC, Universit\'e Paris-Diderot,
PSL Research University, 5 place Jules Janssen, 92195 Meudon, France, \email{pierre.kervella@obspm.fr}.
\and
Institut de Radio-Astronomie Millim\'etrique, 300 rue de la Piscine, 38406 St Martin d'H\`eres, France.
\and
Laboratoire Lagrange (UMR 7293), Universit\'e de Nice-Sophia Antipolis, CNRS, Observatoire de la C\^ote d'Azur, Bd de l'Observatoire, B.P. 4229, F-06304 Nice cedex 4, France.
\and
National Optical Astronomy Observatories, 950 North Cherry Avenue, Tucson, AZ 85719, USA.
\and
European Southern Observatory, Alonso de C{\'o}rdova 3107, Casilla 19001, Santiago 19, Chile.
\and
Instituto de Astronom\'{\i}a, Universidad Cat{\'o}lica del Norte, Avenida Angamos 0610, Antofagasta, Chile.
\and
Universidad de Concepci{\'o}n, Departamento de Astronom\'{\i}a, Casilla 160-C, Concepci{\'o}n, Chile.
}
\date{Received ; Accepted}
\abstract
{
The bright southern star L$_2$\,Pup is a particularly prominent asymptotic giant branch (AGB) star, located at a distance of only 64\,pc.
We report new adaptive optics observations of L$_2$\,Pup at visible wavelengths with the SPHERE/ZIMPOL instrument of the VLT that confirm the presence of the circumstellar dust disk discovered recently.
This disk is seen almost almost edge-on at an inclination of 82$^\circ$.
The signature of its three-dimensional structure is clearly observed in the map of the degree of linear polarization $p_L$.
We identify the inner rim of the disk through its polarimetric signature at a radius of 6\,AU from the AGB star.
The ZIMPOL intensity images in the $V$ and $R$ bands also reveal a close-in secondary source at a projected separation of 2\,AU from the primary. Identification of the spectral type of this companion is uncertain due to the strong reddening from the disk, but its photometry suggests that it is a late K giant with comparable mass to the AGB star.
We present refined physical parameters for the dust disk derived using the RADMC-3D radiative transfer code. We also interpret the $p_L$ map using a simple polarization model to infer the three-dimensional structure of the envelope.
Interactions between the inner binary system and the disk apparently form spiral structures that propagate along the orthogonal axis to the disk to form streamers. Two dust plumes propagating orthogonally to the disk are also detected. They originate in the inner stellar system and are possibly related to the interaction of the wind of the two stars with the material in the disk.
Based on the morphology of the envelope of L$_2$\,Pup, we propose that this star is at an early stage in the formation of a bipolar planetary nebula.
}

\maketitle

\newpage


\section{Introduction}

Evolved stars are important contributors to the enrichment of heavy elements in the interstellar medium and, more generally, to the chemical evolution of the Universe. Since they are much more numerous than high mass stars, the low and intermediate mass stars ($1-8\,M_\odot$) play a particularly important role.
L$_2$\,Puppis (\object{HD 56096}, \object{HIP 34922}, \object{HR 2748}) is an asymptotic giant branch (AGB) Mira-like variable with an uncertain mass estimated at $0.7\,M_\odot$ by \citetads{lykou15}, $1.7\,M_\odot$ by \citetads{1998NewA....3..137D} and $2\,M_\odot$ by \citetads{2014A&A...564A..88K}. This is probably the second nearest known AGB star at $d = 64$\,pc \citepads[$\pi = 15.61 \pm 0.99$\,mas,][]{2007A&A...474..653V}, i.e.~only 9\,pc farther than R\,Dor and 30\% closer than Mira. It is also one of the brightest at $m_V \approx 7$ and $m_K \approx -1.5$. L$_2$\,Pup is therefore a particularly prominent benchmark object to monitor the final stages of the evolution of moderately massive stars.

In 2013, \citetads{2014A&A...564A..88K} obtained a series of NACO images in 12 narrow-band filters covering $\lambda=1.0-4.0\,\mu$m. They revealed a dust lane in front of the star that exhibits a high opacity in the $J$ band and becomes translucent in the $H$ and $K$ bands. In the $L$ band, thermal emission from the dust disk was also detected, as well as a large loop that extends to more than 10\,AU. New observations by \citetads{lykou15} using the sparse aperture masking mode of NACO confirmed the observed structures.
From aperture-synthesis imaging combining NACO and VLTI/AMBER, \citetads{2015ohnaka} identified clumpy dust clouds and also confirm the presence of an elongated central emission in the east-west direction. They also reveal that the southern half of the central star is severely obscured by the nearly edge-on disk.
The hypothesis proposed by \citetads{2014A&A...564A..88K} is that the circumstellar dust of L$_2$\,Pup is distributed in a disk seen at an almost edge-on inclination of $84^\circ$.
This disk may act as a collimator for the AGB star mass loss. However, the true three-dimensional (3D) geometry of the envelope was not constrained by the NACO infrared images.
In addition, no companion was detected in the NACO images, although the large dust loop observed in the $L$ band indicated the possible presence of a secondary star.

In the present work, we report diffraction limited polarimetric images at visible wavelengths from the SPHERE/ZIMPOL adaptive optics system (Sect.~\ref{observations}). The resulting intensity and degree of linear polarization maps allow us to address two questions: (1) What is the 3D geometry of the dust distribution around L$_2$\,Pup? And (2) is L$_2$\,Pup a binary star? After an analysis of the images (Sect.~\ref{analysis}), we discuss the implications in Sect.~\ref{discussion}.

\section{Observations and data reduction}\label{observations}

\subsection{Observations\label{observ}}

The Spectro-Polarimetric High-contrast Exoplanet REsearch \citepads[SPHERE, ][]{2008SPIE.7014E..18B} is a high performance adaptive optics system \citepads{2014SPIE.9148E..1UF} installed at the Nasmyth focus of the Unit Telescope~3 of the Very Large Telescope. We observed L$_2$\,Pup on the night of 7 December 2014 using SPHERE and the Zurich IMaging POLarimeter \citepads[ZIMPOL, ][]{2014SPIE.9147E..3WR} during the Science Verification of the instrument.
We observed L$_2$\,Pup and a point spread function (PSF) calibrator, $\beta$\,Col (\object{HD 39425}, spectral type K1III) in the polarimetric P2 mode. $\beta$\,Col also acts as a photometric calibrator as it belongs to the catalogs by \citetads{2002A&A...393..183B} and \citetads{1999AJ....117.1864C}. The ZIMPOL instrument includes two cameras that allow for simultaneous observations at two different wavelengths. We used the $V$ (for camera 1, hereafter referred to as {\tt cam1}) and $N_R$ ({\tt cam2}) filters, that have respective central wavelengths of $\lambda_V = 554.0$ and $\lambda_R = 645.9$\,nm, and full width half maximum bandwidths of 80.6 and 56.7\,nm.
The list of observations is given in Table~\ref{obs_log}. The observations of $L_2$\,Pup were taken both with and without Lyot coronagraph ({\tt V\_CLC\_S\_WF}). The PSF reference $\beta$\,Col was observed only without coronagraph, and through the {\tt ND2} neutral density filter, that has a photometric transmission of 0.61\% in $V$ and 0.76\% in $N_R$\footnote{According to the SPHERE User Manual version P95.2.}.

\begin{figure}[]
        \centering
        \includegraphics[width=7cm]{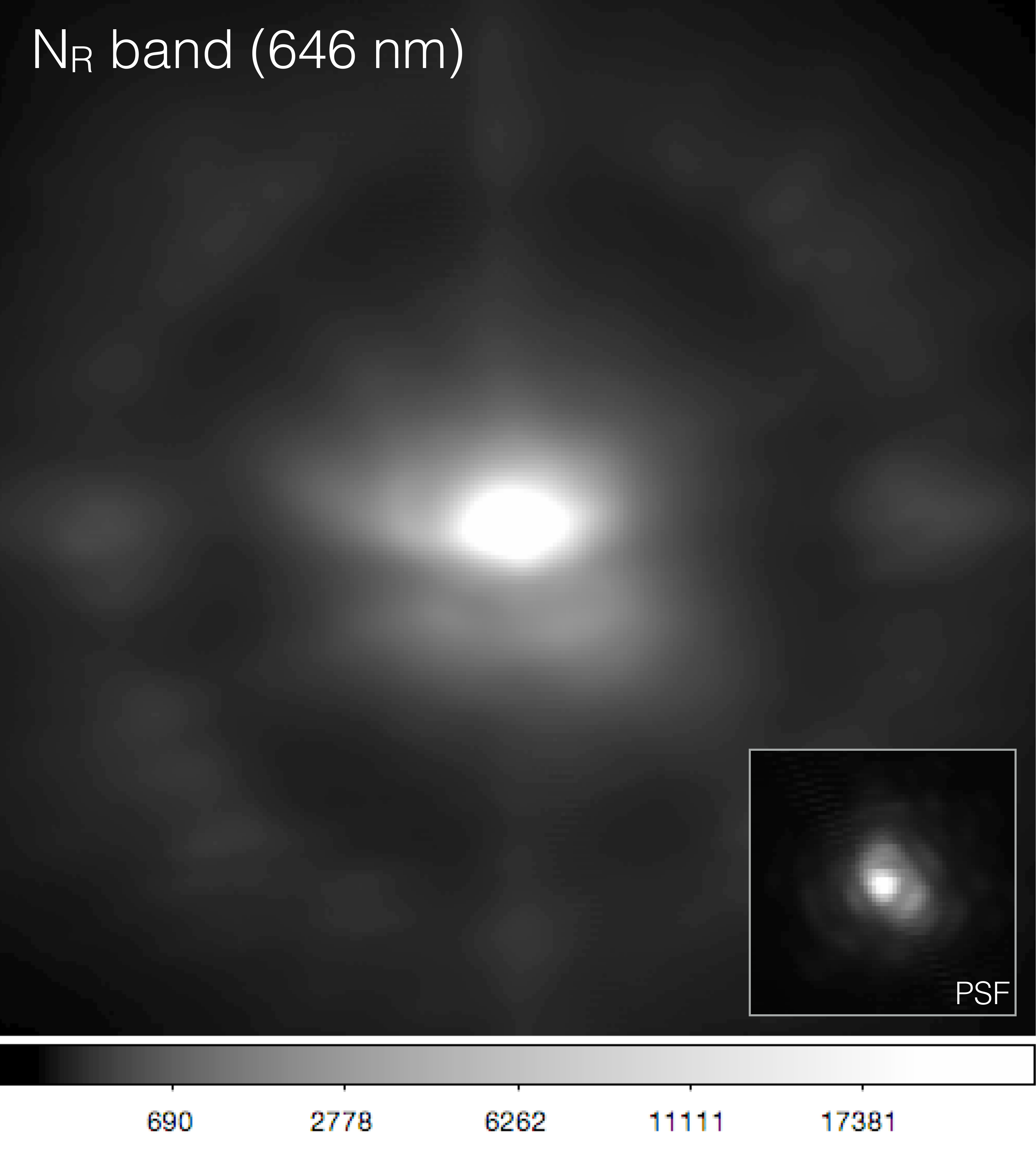}
        \caption{Non-deconvolved intensity image of L$_2$\,Pup in the $N_R$ band, with a square root intensity scale (in ADU) and a field of view of $0.91\arcsec \times 0.91\arcsec$. The PSF ($\beta$\,Col) is shown as an insert at the same scale.
        \label{nondeconv}}
\end{figure}

\begin{figure}[]
        \centering
        \includegraphics[width=3.7cm]{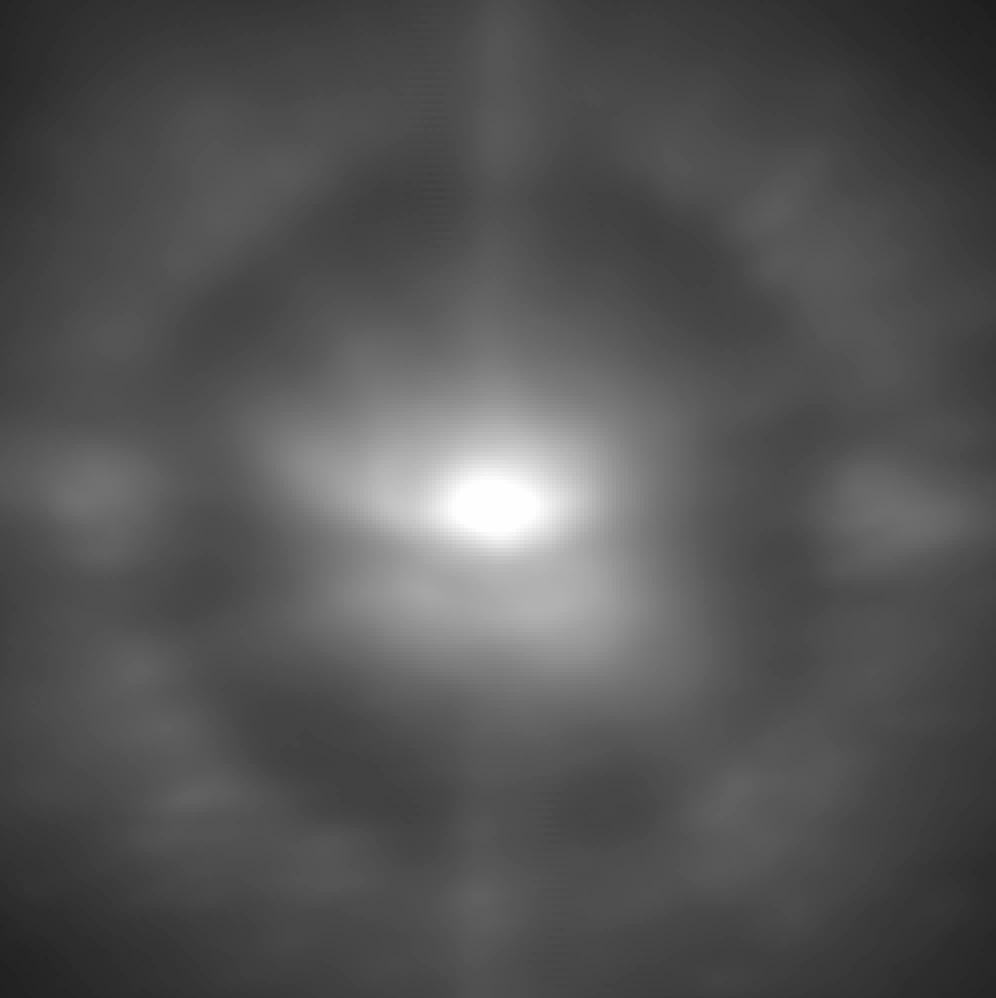} \hspace{2mm}
        \includegraphics[width=3.7cm]{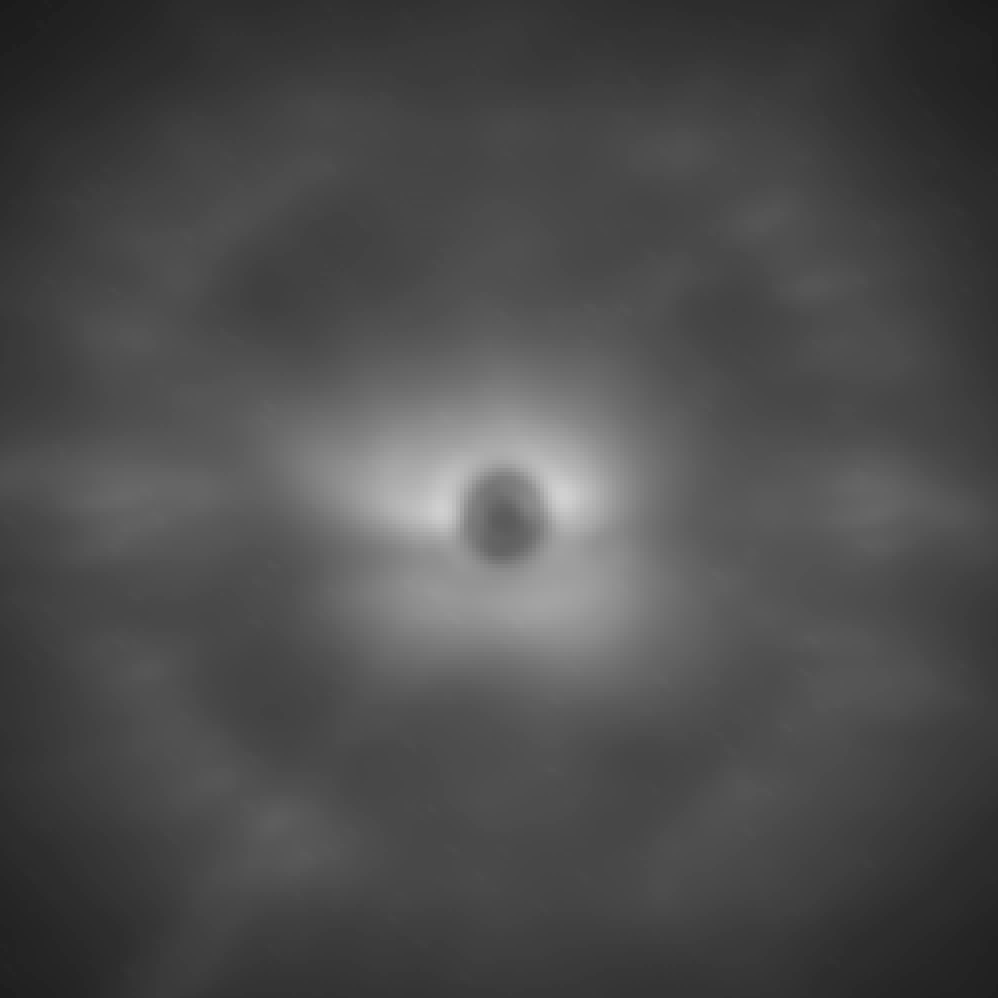}
        \includegraphics[width=3.7cm]{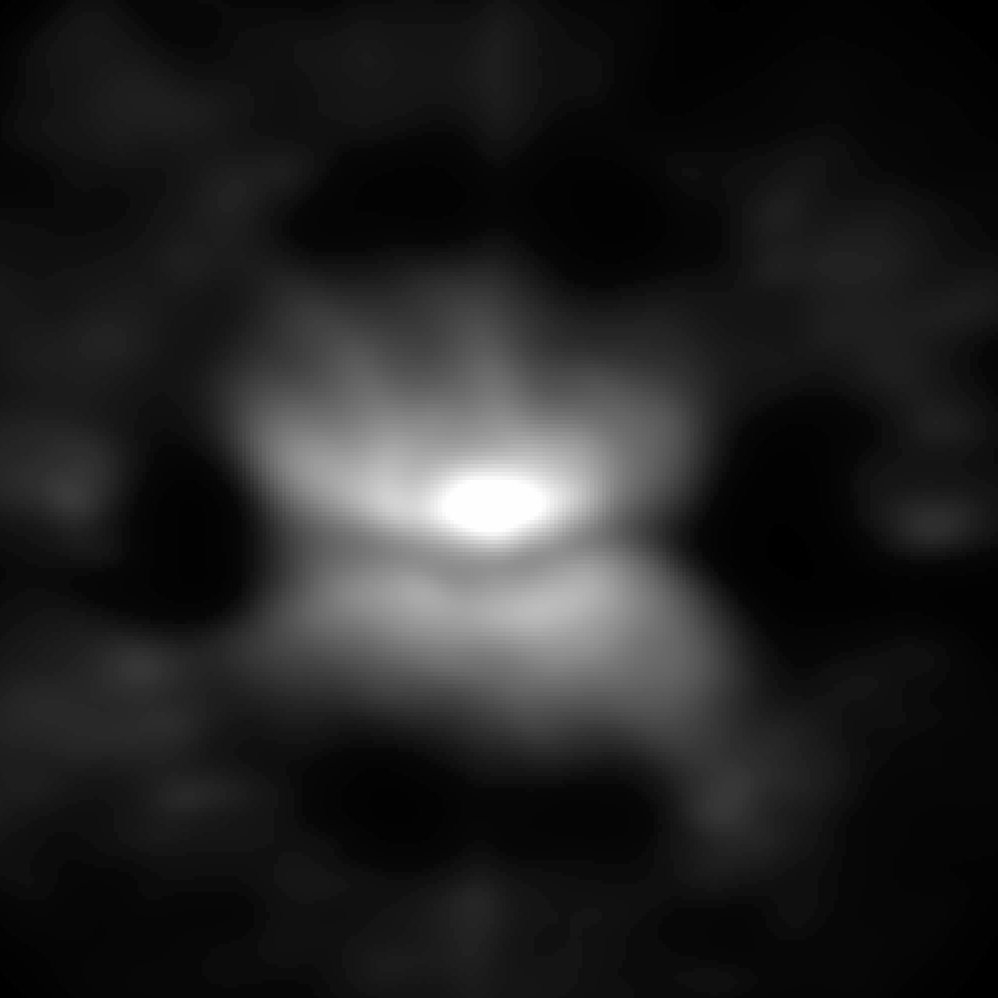} \hspace{2mm}
        \includegraphics[width=3.7cm]{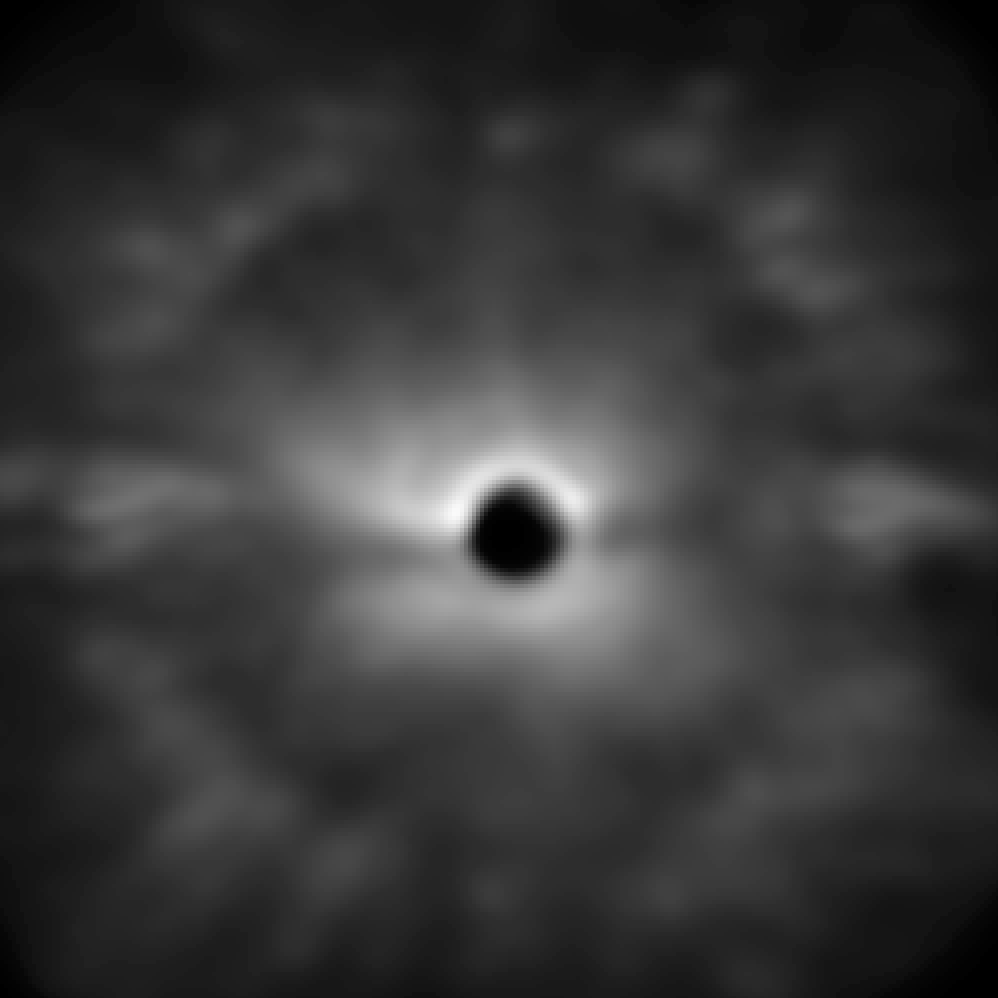}
        \caption{{\it Top:} Comparison of the image of L$_2$\,Pup in the $N_R$ band without coronagraph (left) and the image in the same filter with coronagraph (right). {\it Bottom:} Same images after deconvolution (20 steps of IRAF's \texttt{lucy} algorithm). The logarithmic color scales have been matched taking into account the different exposure times. The field of view is $0.91\arcsec \times 0.91\arcsec$.
        \label{coro-nocoro}}
\end{figure}

We processed the raw cubes using the ZIMPOL data reduction pipeline, in its pre-release version 0.14.0\footnote{Downloadable from \url{ftp://ftp.eso.org/pub/dfs/pipelines/sphere/}}. The observations resulted in 24 data cubes for observations \#1 and \#2 (Table~\ref{obs_log}), containing the frames recorded with the two cameras, and 48 cubes for observation \#3. Each of these cubes was processed separately using the pipeline to produce Stokes $+Q$, $-Q$, $+U$ and $-U$ frames for each of the two cameras, together with their associated intensity frames $I_Q$ and $I_U$. The resulting average frames were then precisely aligned and derotated using custom {\tt python} routines. We observed a fixed rotation offset of 3$^\circ$ clockwise of {\tt cam2} images relative to {\tt cam1} taken as the reference in the following. The vertical axis of {\tt cam1} is oriented north-south, with north up and east to the left (all images are oriented following this convention in the present work). Examples of the resulting images of L$_2$\,Pup and the PSF calibrator are presented in Fig.~\ref{nondeconv}.
The images obtained with the Lyot coronagraph did not show a significant improvement in the detection of the faint extensions of the disk, while the presence of the occulter created perturbations to the wings of the PSF.
As a result, the coronagraphic intensity images are poorly suited for deconvolution (Fig.~\ref{coro-nocoro}), and we did not consider them further in this work. Fortunately, the polarimetric parameters of the coronagraphic observations are of excellent quality, as the PSF wing signature is naturally subtracted during the computation of the Stokes Q and U maps. We will therefore consider the coronagraphic $p_L$ maps together with the non-coronagraphic versions in order to check for the stability of the polarimetric measurement.

\begin{table*}
        \caption{Log of the SPHERE/ZIMPOL observations of L$_2$ Pup and its associated PSF calibrator, $\beta$\,Col.}
        \centering          
        \label{obs_log}
        \begin{tabular}{lllcccccccc}
	\hline\hline
        \noalign{\smallskip}
        \# & MJD & Star & Coronagraph & DIT & NDIT & Dither & Polar & Total exp. & Seeing & AM \\
	 & & & & [s] & & & cycles & [s] & [''] & \\
         \hline         
        \noalign{\smallskip}
	1 & 56998.3024 & $\beta$\,Col & No & 1.2 & $16 \times 2$ & 6 & 1 & 230.4 & 0.74 & 1.11 \\
	2 & 56998.3229 & L$_2$\,Pup & No & 1.2 & $50 \times 2$ & 6 & 1 & 720 & 0.77 & 1.08 \\
	3 & 56998.3558 & L$_2$\,Pup & Yes & 12 & $4 \times 2$ & 6 & 2 & 1152 & 0.67 & 1.14 \\
        \hline
        \end{tabular}
        \tablefoot{MJD is the average modified julian date of the exposures. DIT is the individual exposure time of the SPHERE frames, and NDIT is the total number of frames recorded. Dither is the number of dithering positions, and AM is the airmass.}
\end{table*}

\subsection{Polarimetric quantities and deconvolution}

From the average $+Q$, $-Q$, $+U$, $-U$, $I_Q$ and $I_U$ frames, we derive the degree of linear polarization $p_L$, the polarization electric-vector position angle $\theta$ and the intensity map $I$ using:
\begin{equation}
q = \frac{[+Q] - [-Q]}{2\,I_Q}, \hspace{5mm} u = \frac{[+U] - [-U]}{2\,I_U}
\end{equation}

\begin{equation}
p_L = \sqrt{ q^2 + u^2 }, \hspace{5mm} \theta = \arctan \left( \frac{q}{u} \right),  \hspace{5mm} I = \frac{I_Q + I_U}{2}.
\end{equation}

\begin{figure}[]
        \centering
        \frame{\includegraphics[width=3.7cm]{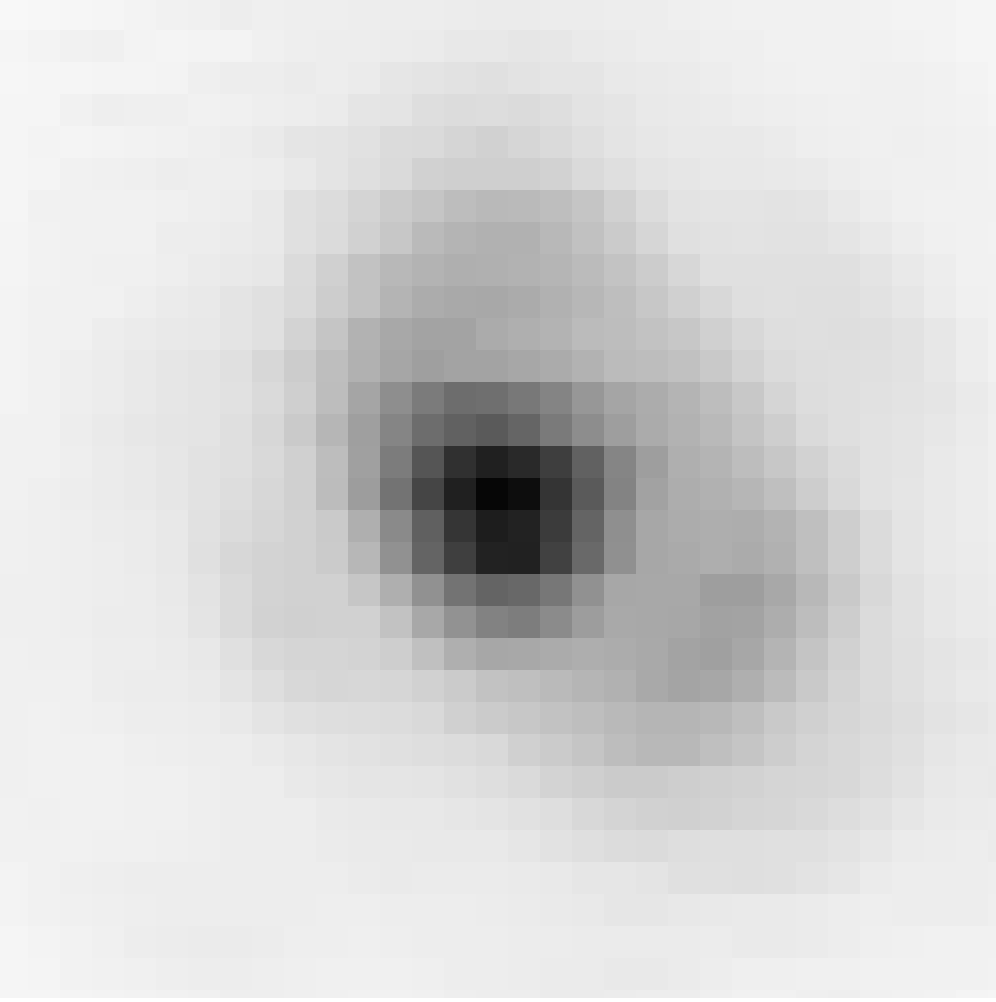}} \hspace{2mm}
        \frame{\includegraphics[width=3.7cm]{Figures/iq-nr-psf.png}}
        \caption{Comparison of the $I_Q$ (left) and $I_U$ (right) images of the PSF star $\beta$\,Col. The gray scale is identical between the two images, and the field of view is $0.113\arcsec \times 0.113\arcsec$.
        \label{iq-iu-psf}}
\end{figure}

\begin{figure*}[]
        \centering
        
        \includegraphics[width=7cm]{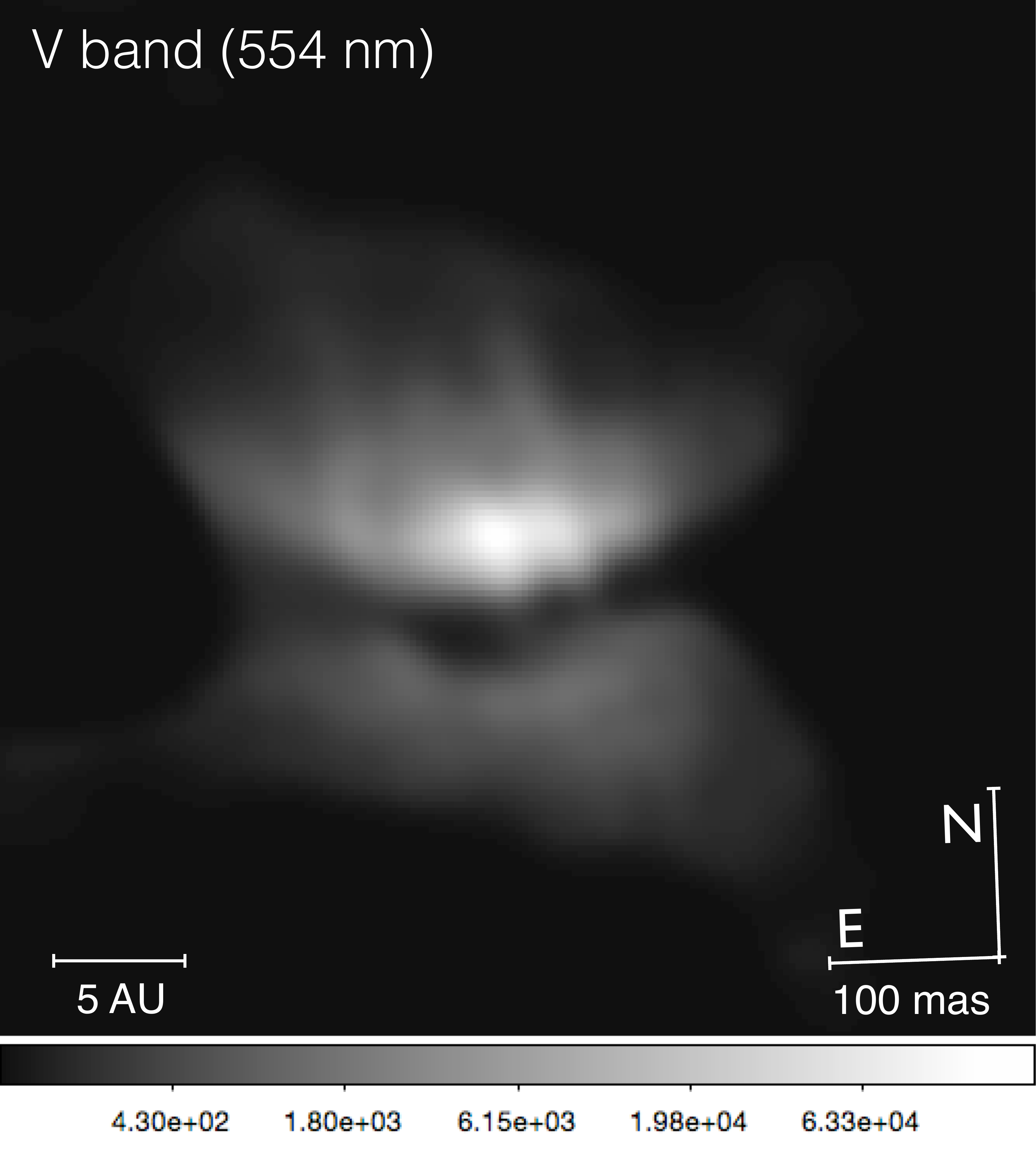} \hspace{5mm} \includegraphics[width=7cm]{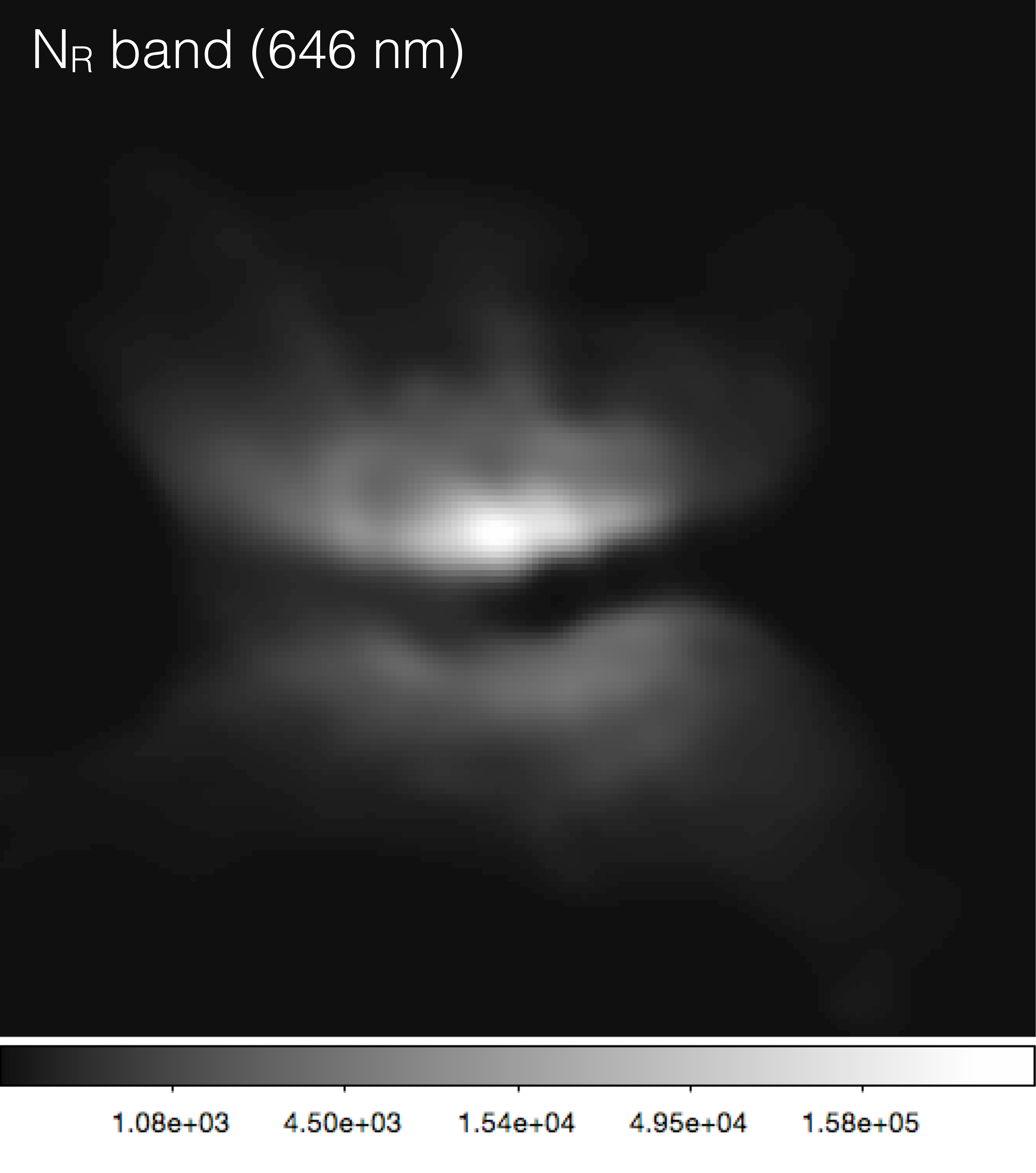}\\
        \includegraphics[width=7cm]{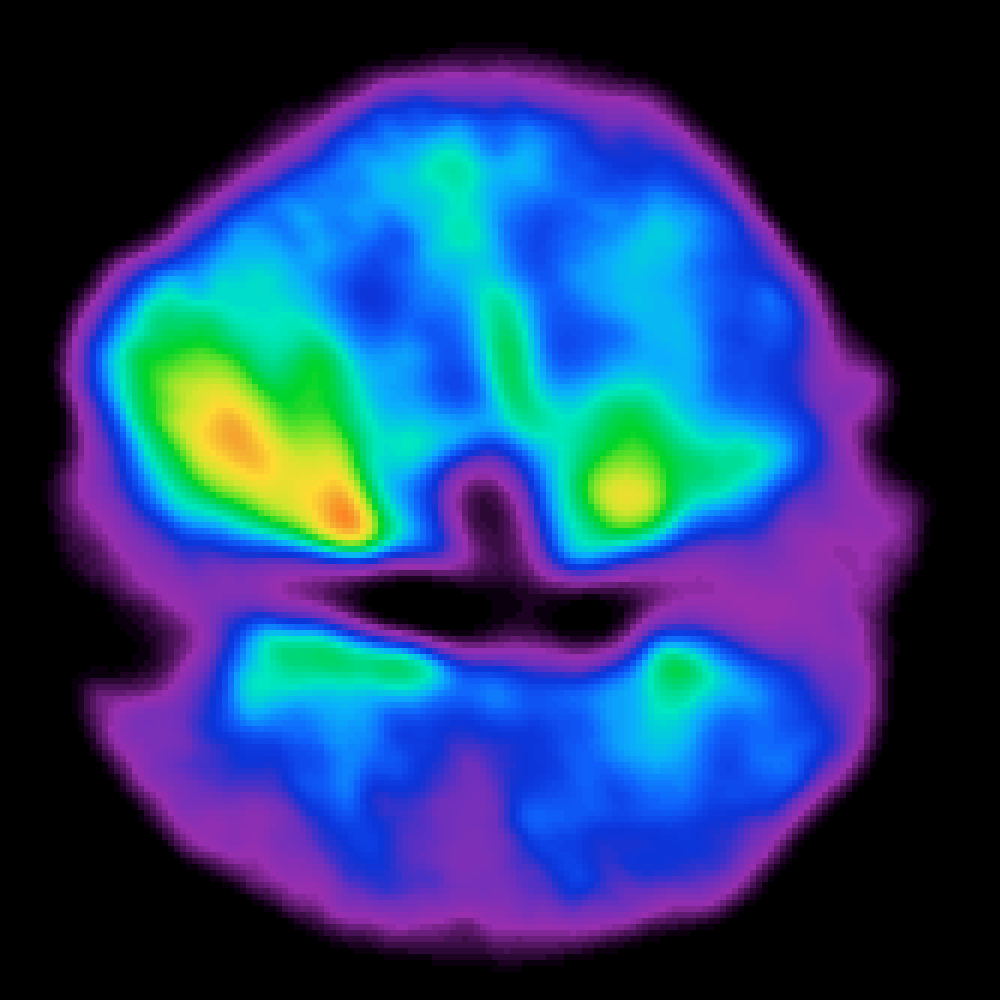} \hspace{5mm} \includegraphics[width=7cm]{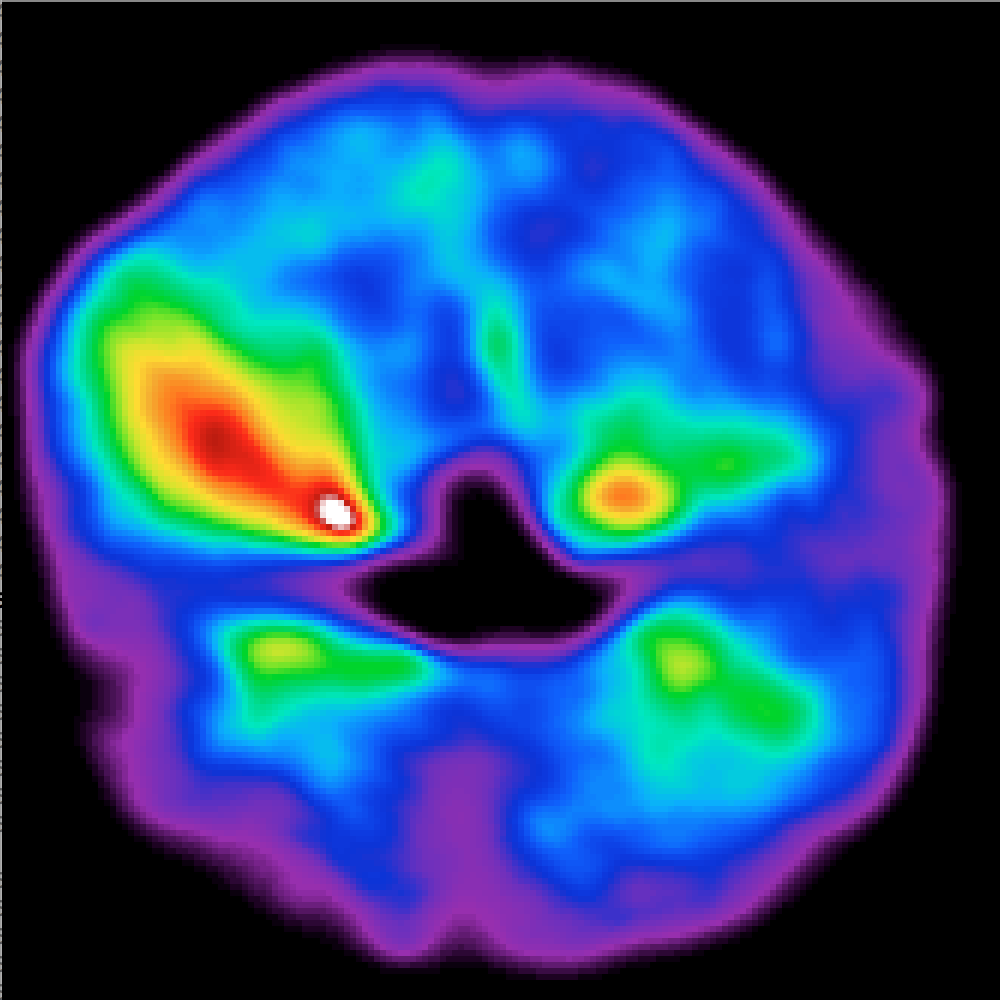}\\
        \includegraphics[width=7cm]{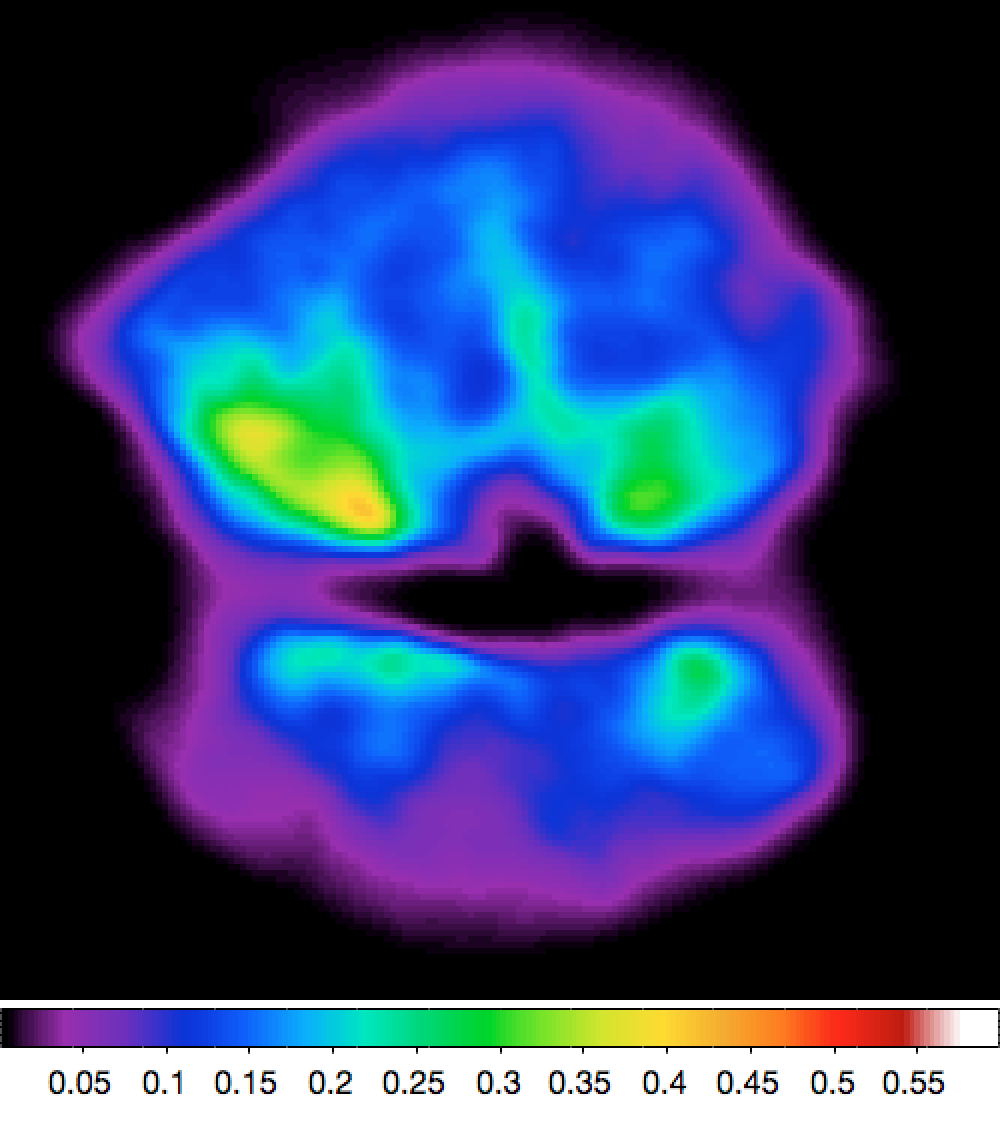} \hspace{5mm} \includegraphics[width=7cm]{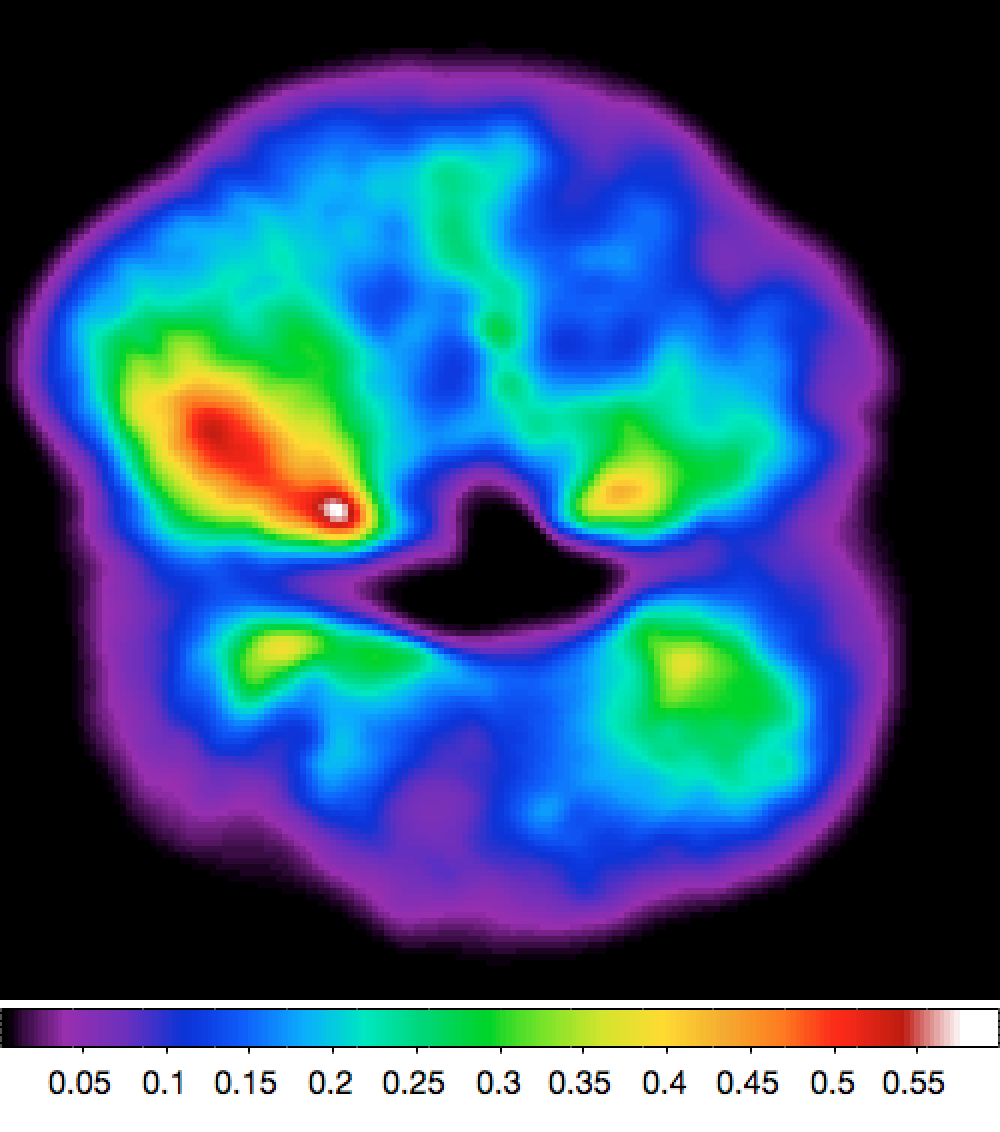}
      

        \caption{
        {\it Top:} Deconvolved intensity images of L$_2$\,Pup in the $V$ (left) and $N_R$ (right) bands (logarithmic intensity scale, in ADU).
        {\it Middle:} Deconvolved map of the degree of linear polarization  $p_L$ in the $V$ (left) and $N_R$ (right) bands from the non-coronagraphic observations (linear intensity scale).
        {\it Bottom:} $p_L$ in the $V$ (left) and $N_R$ (right) bands from the coronagraphic frames (linear intensity scale).
        The field of view is $0.60\arcsec \times 0.60\arcsec$.
        \label{deconv80}}
\end{figure*}

\noindent
We deconvolved the intensity and $p_L$ images of L$_2$\,Pup using the $\beta$\,Col intensity maps $I$ as the dirty beams, and the Lucy-Richardson (L-R) algorithm implemented in the IRAF\footnote{\url{http://iraf.noao.edu}} software package. 
We checked that the small differences between the $I_Q$ and $I_U$ images of the PSF star (Fig.~\ref{iq-iu-psf}) do not introduce artefacts in the deconvolution process when using the average intensity image $I$ as the dirty beam.
\citetads{2005ApJ...627..701M} showed that the Lucy-Richardson algorithm preserves the photometric accuracy of the original images relatively well compared to other classical algorithms.
We stopped the L-R deconvolution after 80 iterations (uniformly for both cameras), as the deconvolved images do not show a significant evolution for additional processing steps. The resulting $I$ and $p_L$ deconvolved frames are presented in Fig.~\ref{deconv80}.
The quality of the AO correction and the stability of the atmospheric seeing during the observations (Table~\ref{obs_log}) results in a very good aptitude of the $L_2$\,Pup images for deconvolution. On the PSF calibrator $\beta$\,Col, we measure a Strehl ratio $S \approx 30$\% in the $V$ filter, and $S=40$ to $45$\% in the $N_R$ filter. This is equivalent to the Strehl ratios obtained with NACO respectively in the $J$ and $H$ bands, in good atmospheric conditions. The measured full width at half maximum (FWHM) of the PSF image is approximately $16.5 \times 18$\,mas in the $V$ band and $17 \times 20$\,mas in the $N_R$ band.

The $p_L$ maps show clearly the presence of four maxima located on the side, below and above the dark dust lane. This behavior is typical of light scattering in a thick circumstellar dust disk, as the light is scattered and transmitted on both sides of the disk, with a maximum polarization for a scattering angle close to $90^\circ$ (see Sect.~\ref{polmodel} for a further discussion).

\subsection{Pixel scale and field orientation}

We aligned the {\tt cam2} images on the orientation of {\tt cam1} by rotating them by $3\,\deg$ counter clockwise, as described in Sect.~\ref{observ}. 
We adopt a pixel scale of $3.628 \pm 0.036$\,mas\,pix$^{-1}$ and a position angle of the vertical axis with respect to north of $357.95 \pm 0.55\,\deg$ for both {\tt cam1} and {\tt cam2}.
 These parameters were derived from commissioning observations of several astrometric reference sources by Ch. Ginski (private communication), including five astrometric binaries with high quality orbital solutions and the quadruple system $\theta^1$\,Ori\,B.

\section{Analysis}\label{analysis}

\subsection{Photometry and astrometry\label{photometry-astrometry}}

From the ZIMPOL images, the total flux of L$_2$\,Pup and its close envelope (within $0.5\arcsec$ of the central source) relative to $\beta$\,Col is estimated to $f_{\mathrm{L}_2\,\mathrm{Pup}}/f_{\beta\,\mathrm{Col}} = 0.027 \pm 0.001$ in the $V$ band and $f_{\mathrm{L}_2\,\mathrm{Pup}}/f_{\beta\,\mathrm{Col}} = 0.049 \pm 0.002$ in the $N_R$ band. Although ZIMPOL's $N_R$ filter has a narrower bandpass than the standard $R$ filter, we consider here that they are equivalent. The $V$ and $R$ magnitudes of $\beta$\,Col were taken from \citetads{1978A&AS...34..477M}.

\begin{table}
        \caption{Photometry of L$_2$ Pup and its companion.}
        \centering          
        \label{photom_table}
        \begin{tabular}{lrr}
	\hline\hline
         Star & $m_V$ & $m_R$ \\
         \hline         
        \noalign{\smallskip}
         $\beta$\,Col (calibrator) & $3.12 \pm 0.01$ & $2.27 \pm 0.01$ \\
	\hline
        \noalign{\smallskip}
         L$_2$\,Pup (total) & $7.03 \pm 0.05$ & $5.54 \pm 0.05$ \\
	Disk only & $7.36 \pm 0.10$ & $5.97 \pm 0.10$ \\
         L$_2$\,Pup~A & $8.72 \pm 0.10$ & $6.93 \pm 0.10$ \\
         L$_2$\,Pup~B & $10.29 \pm 0.10$ & $8.75 \pm 0.10$ \\
        \hline
        \end{tabular}
\end{table}

\begin{figure}[]
        \centering
        \frame{\includegraphics[width=4.2cm]{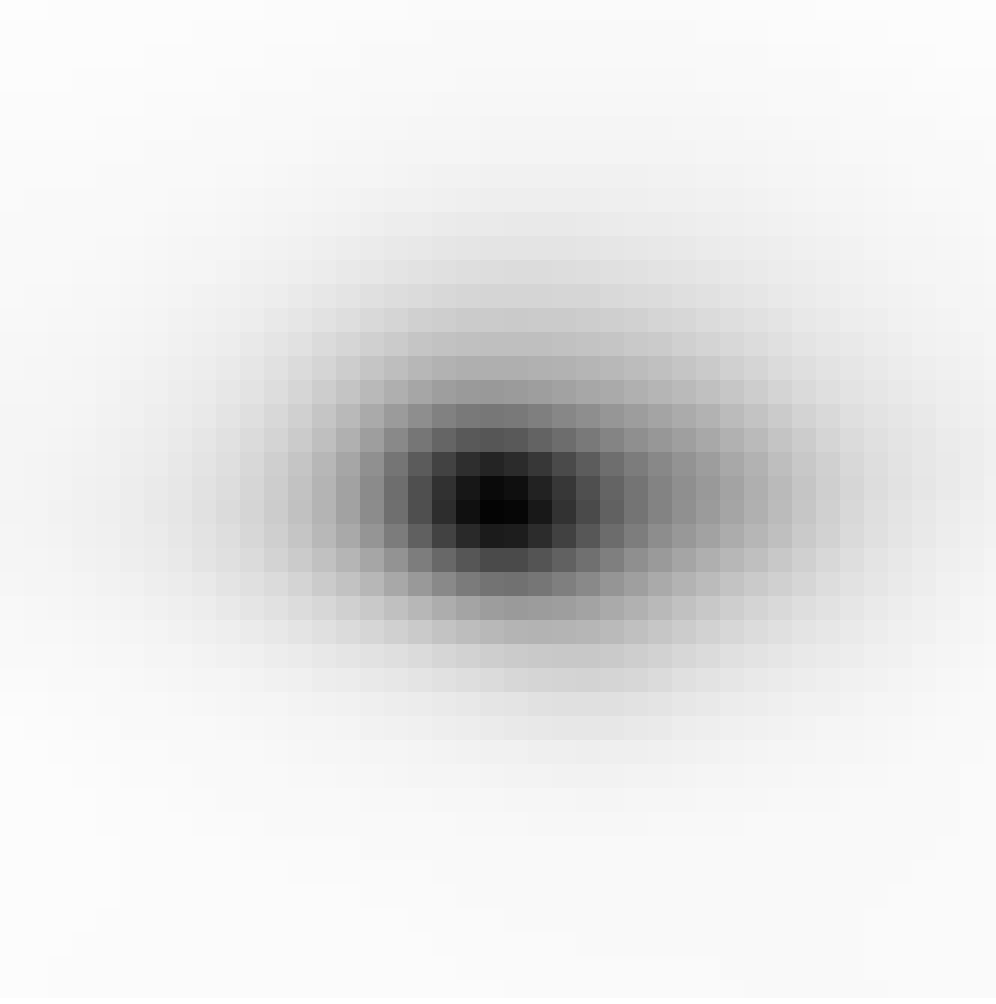}} \hspace{2mm}
        \frame{\includegraphics[width=4.2cm]{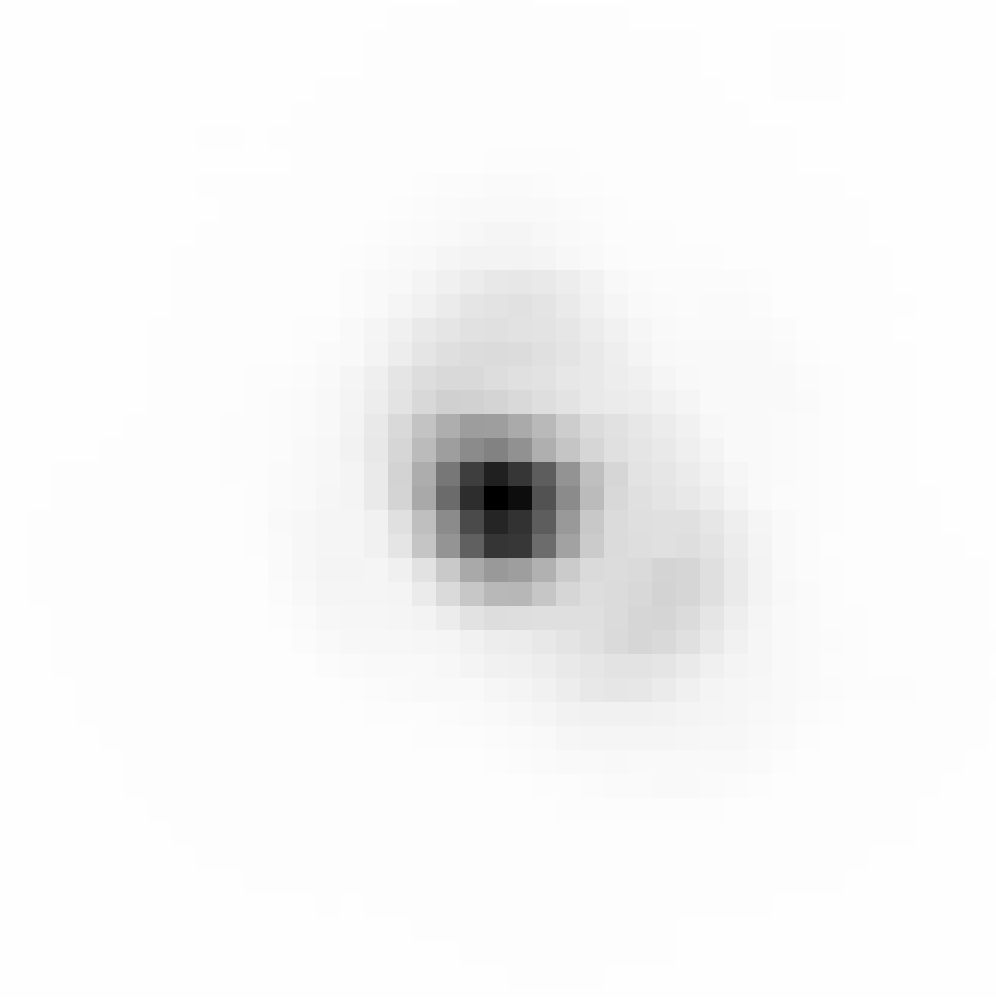}}
        \caption{Non-deconvolved intensity images in $N_R$ of the central source of L$_2$\,Pup (left) and the PSF calibrator $\beta$\,Col (right). The gray scale is linear between the minimum and maximum value of each image, and the field of view is $0.151\arcsec \times 0.151\arcsec$.
        \label{nondec}}
\end{figure}

\begin{figure}[]
        \centering
        \includegraphics[width=4.2cm]{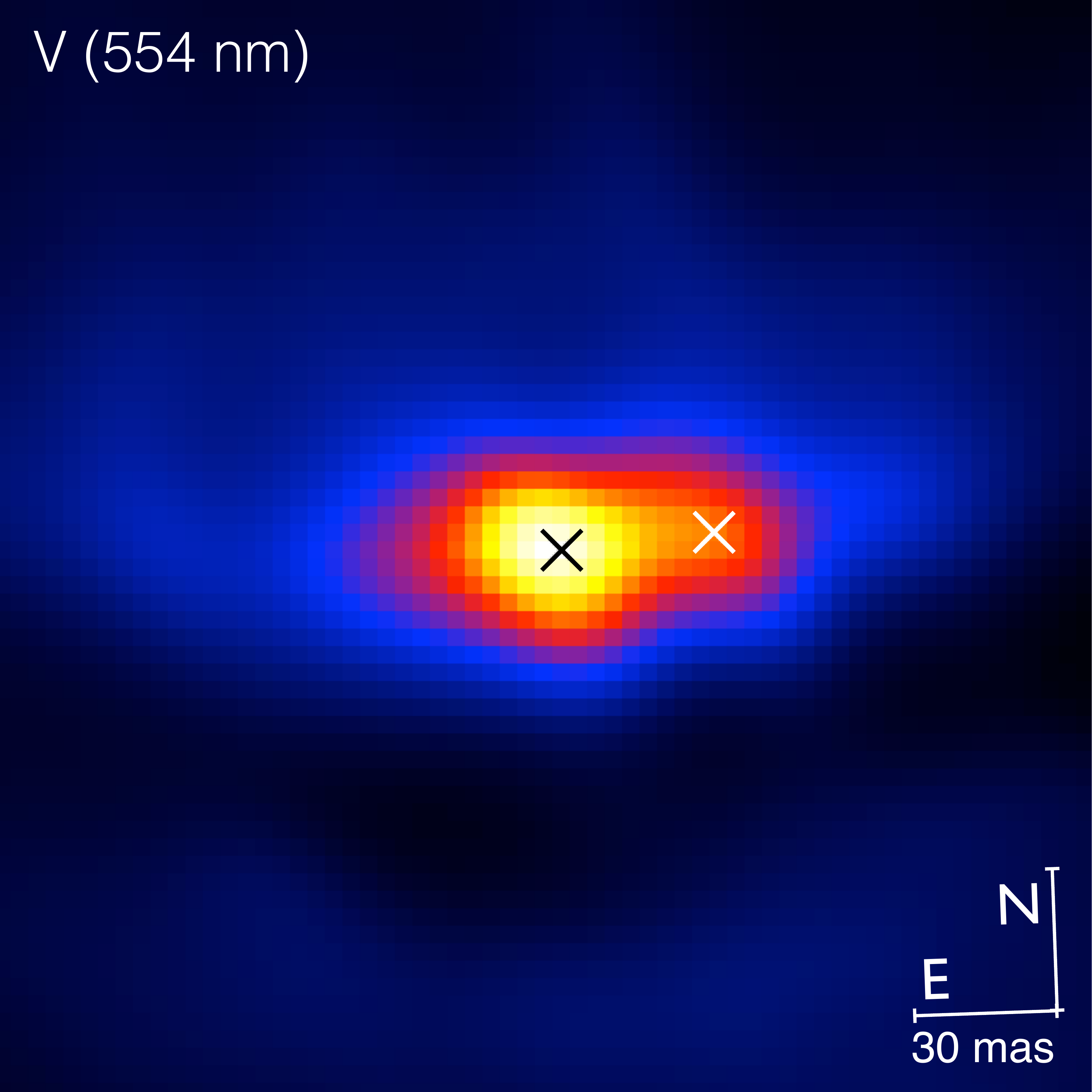}\hspace{2mm}
        \includegraphics[width=4.2cm]{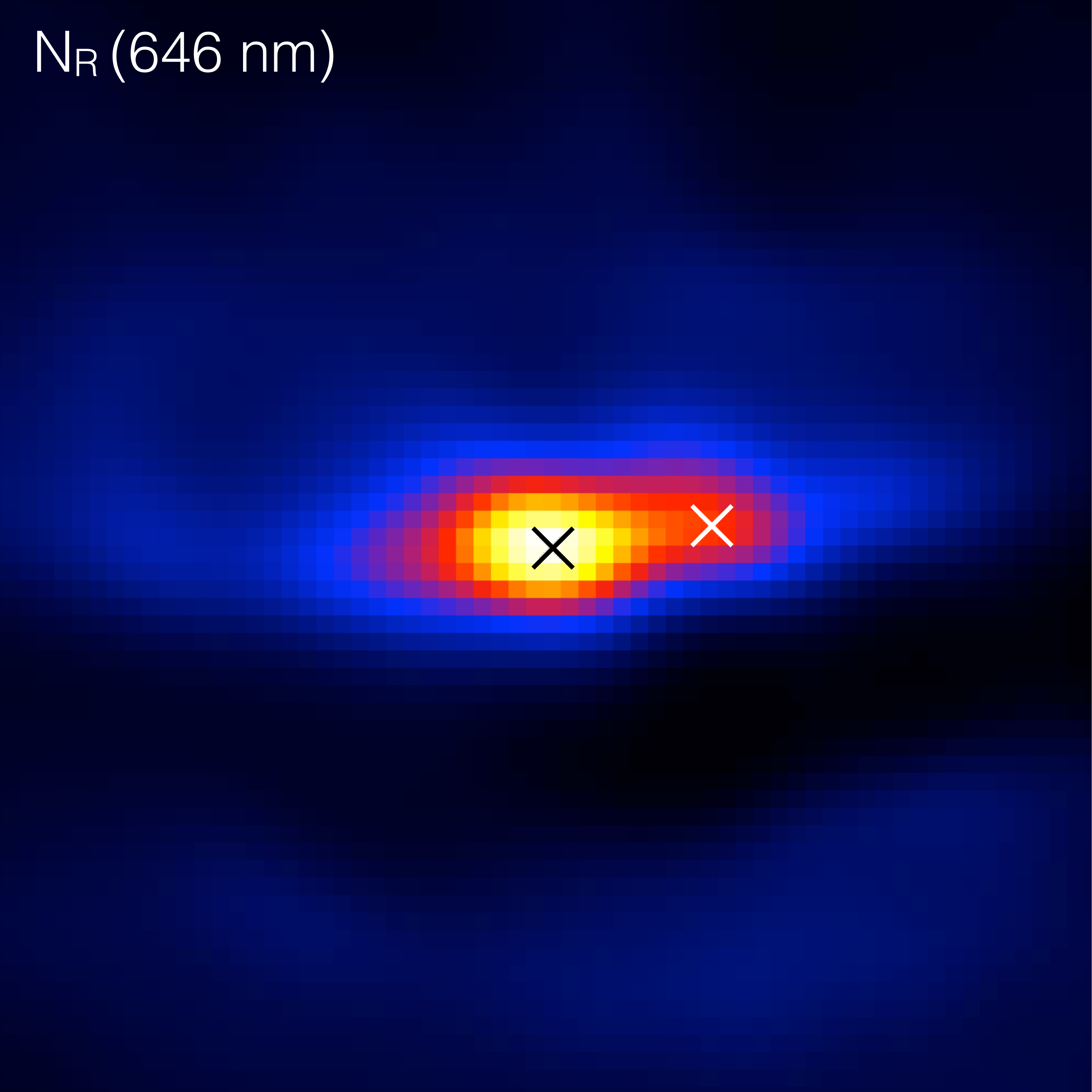}
        \caption{Position of L$_2$\,Pup and its companion in the deconvolved $V$ (left) and $N_R$ (right) intensity images. The positions of the two stars are marked with crosses. The field of view is $0.227\arcsec \times 0.227\arcsec$.
        \label{companion}}
\end{figure}

The central source appears noticeably elongated in the ZIMPOL intensity images (Fig.~\ref{nondec}), and after deconvolution, a secondary source is detected, as shown in Fig.~\ref{companion}.
The astrometric position of the fainter source relative to the central AGB star in both the $V$ and $N_R$ images is found to be: $\left[ \Delta x, \Delta y\right] = \left[ +9\ \pm 1\,\mathrm{pix}, +1 \pm 1\,\mathrm{pix}\right]$, corresponding to $\left[ \Delta \alpha, \Delta \delta \right] = \left[ 32.7 \pm 3.6\,\mathrm{mas}, -3.6 \pm 3.6\,\mathrm{mas} \right]$, and a projected separation of $32.9 \pm 3.6$\,mas, or $2.1 \pm 0.2$\,AU considering the \emph{Hipparcos} parallax.
Due to the high geometrical inclination and thickness of the disk, it is likely that part of the photosphere of the AGB star is obscured \citepads{2014A&A...564A..88K}. This effect could bias the apparent astrometric position, as well as that of the companion. However, considering that the vector direction between the AGB star and its companion is mostly aligned with the major axis of the disk, the relative position of the two objects should only be mildly affected.

\begin{figure}[]
        \centering
        \includegraphics[width=7cm]{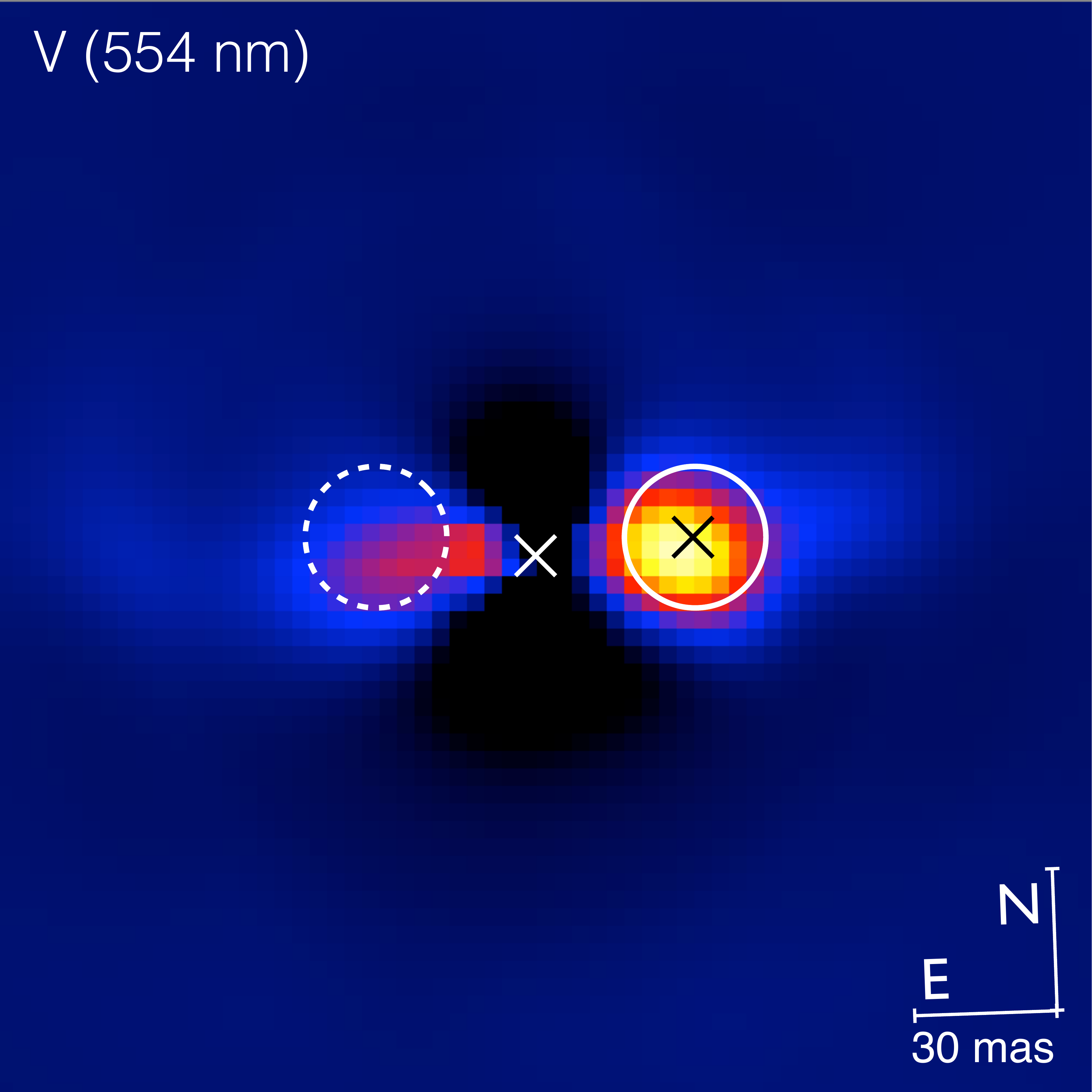}
        \caption{Radial median subtracted image of L$_2$\,Pup's companion in the $V$ band. The photometry of L$_2$\,Pup~B is computed from the integration of the flux enclosed in the solid circle, while the background is estimated in the dashed circle. The positions of the two stars are marked with crosses, and the field of view is $0.227\arcsec \times 0.227\arcsec$.
        \label{companion_photometry}}
\end{figure}

The photometry of the different parts of the L$_2$\,Pup system is listed in Table~\ref{photom_table}. The majority of the flux in the $V$ and $N_R$ bands is scattered light coming from the disk, as the contribution of the two stars to the total flux of the system reaches only 26\% in $V$ and 32\% in $N_R$.
To isolate the photometric contribution of the secondary source L$_2$\,Pup~B, we first subtracted the radial median profile (centered on the primary star) from the deconvolved images to remove the flux contribution of the background. We then measured the integrated flux over a disk with a radius of 4 pixels, as shown in Fig.~\ref{companion_photometry}, and we subtracted the background flux measured on the opposite side of the central star. This second background subtraction is intended to remove the scattered light contribution by the disk.
To estimate the flux from L$_2$\,Pup~A, we adopted the same approach, but we considered the deconvolved image without radial median subtraction, and we subtracted the background from the same background region as for L$_2$\,Pup~B (Fig.~\ref{companion_photometry}). We obtain a flux ratio of $f_V(B)/f_V(A) = 24 \pm 10\%$ in the $V$ band and $f_{NR}(B)/f_{NR}(A) = 19 \pm 10\%$ in the $N_R$ band.
The stated error bars are conservative, to take into account the uncertainty in the neutral density filter transmission, deconvolved flux distribution and estimation of the background level, that may not be homogeneous between the eastern and western sides of the disk.
This translates into $\pm 0.1$\,mag uncertainties for the individual components (disk, A and B), and we adopt $\pm 0.05$\,mag error bars for the integrated photometry of the whole system.
 
The two objects L$_2$\,Pup~A and B appear relatively faint and red in color. This is caused by the considerable wavelength dependent absorption of the light as it passes through the disk. Considering the geometrical configuration of the disk and binary system, the reddening should however be similar for the two objects. Therefore, the difference in apparent $V$ magnitude that we measure ($\Delta m_V = 1.6$) should be close to the difference in absolute $V$ magnitude between the two objects. A discussion of the properties of L$_2$\,Pup~B is presented in Sect.~\ref{companion_properties}.

\subsection{Polarization model and 3D structure of the disk \label{polmodel}}

To map the cold dust, we take advantage of the high scattering and polarizing efficiency of the dust in the visible domain. The basic principle of our approach is to derive the scattering angle $\theta$ of the light on the dust from the measured degree of linear polarization $p_L$. For that purpose, we have to assume a polarization model $p_L(\theta)$, following the method employed by \citetads{2014A&A...572A...7K} to map the 3D structure of the dusty envelope of the long-period Cepheid RS\,Pup. 
The nature of the dust has an impact on its polarization curve, in particular on the maximum value of $p_L$ (hereafter $p_\mathrm{max}$) and also on the angle of maximum polarization ($\theta_\mathrm{max}$).
The consistently high maximum value of $p_L$ in L$_2$\,Pup's envelope indicates that the dust grain physical properties are probably comparable to those of RS\,Pup. The typical Milky Way dust models exhibit lower maximum values of $p_L$ \citepads[20 to 30\%,][]{2003ApJ...598.1017D}. However, laboratory experiments by \citetads{2007A&A...470..377V} show very high degrees of linear polarization (up to $p_L \approx 50\%$ to 100\% in the visible) on light scattered by fluffy dust grains \citepads[see also][]{2009ApJ...696.2126S}, which may be present around L$_2$\,Pup.

We observe in the deconvolved polarization maps presented in Fig.~\ref{deconv80} a value of $p_\mathrm{max}(V) = 0.46$ and $p_\mathrm{max}(N_R) = 0.61$ for the non-coronagraphic images. The coronagraphic images present comparable but slightly lower maxima of $p_\mathrm{max}(V) = 0.42$ and $p_\mathrm{max}(N_R) = 0.58$. These figures are of the same order as the $p_\mathrm{max} = 0.52 \pm 0.02$ value observed by \citetads{2014A&A...572A...7K} in the $V$ band in the light scattering nebula surrounding RS\,Pup, as well as by \citetads{2008AJ....135..605S} in the dusty nebula of V838\,Mon ($p_\mathrm{max} \approx 0.50$, also in the $V$ band).

\begin{figure}[]
        \centering
        \includegraphics[width=\hsize]{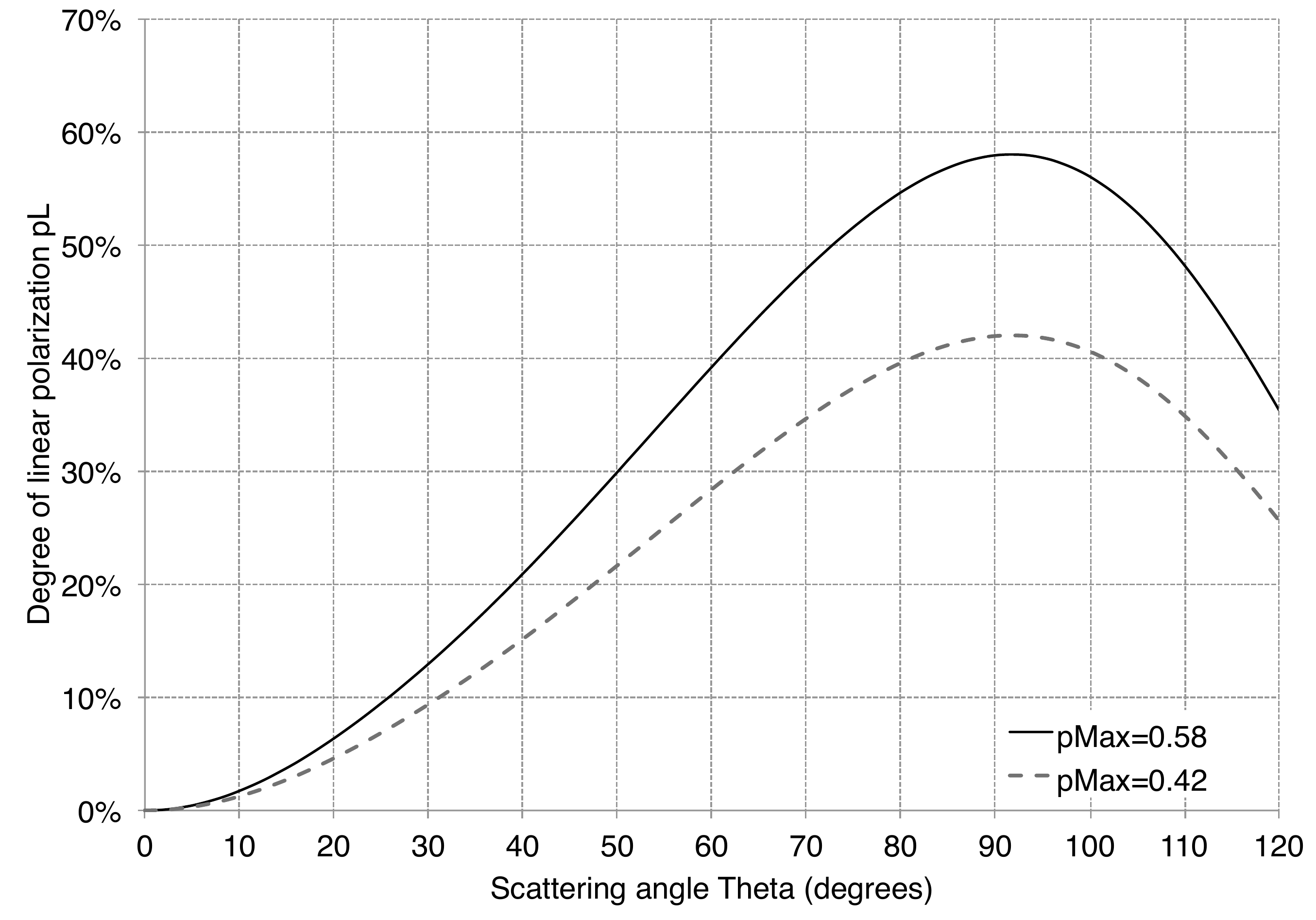}
        \caption{Adopted polarization models $p_L(\theta)$ for the $V$ band (dashed curve) and the $N_R$ band (solid curve).
        \label{polarization-model}}
\end{figure}

For the present analysis of L$_2$\,Pup's polarization, we thus adopt the scaled ``Draine'' polarization model of the Milky Way dust presented by \citetads{2014A&A...572A...7K}, with $p_\mathrm{max}(V)= 0.42 \pm 0.02$ and $p_\mathrm{max}(N_R)= 0.58 \pm 0.02$. The polarization for this model peaks at a scattering angle of $\theta_\mathrm{max}=92^\circ$. The chosen $p_\mathrm{max}$ values were measured on the average $p_L$ maps of the non-coronagraphic and coronagraphic frames, and the error bars cover the range of values between the frames. The statistical errors are significantly smaller, but the deconvolution process introduces an additional uncertainty that we take into account conservatively. The resulting $p_L(\theta)$ model curves are presented in Fig.~\ref{polarization-model}.
The fact that the $p_\mathrm{max}$ value is smaller in the $V$ band than in the $N_R$ band is in agreement with the existing light scattering models. For instance, the Milky Way dust model\footnote{Data available from \url{https://www.astro.princeton.edu/~draine/dust/scat.html}} presented by \citetads{2003ApJ...598.1017D} for a total to selective extinction parameter $R_V = A_V/E(B-V) = 3.1$ gives $p_\mathrm{max} = 0.236$ at $\lambda=547$\,nm, and $p_\mathrm{max} = 0.273$ at $\lambda=649$\,nm. \citetads{2014A&A...572A...7K} presents a more detailed discussion of the choice of polarization model.

In our simple approach, we assume that the scattering angle is lower than or equal to $\theta_\mathrm{max}$, i.e.~that the scattering occurs in the forward direction. The reason for this choice is that at visible wavelengths, forward scattering is far more efficient than backward scattering. This means that the majority of the scattered light we observe around L$_2$\,Pup comes from dust located between us and the AGB star, rather than behind the star. However, for scattering angles close to 90$^\circ$, there is an unresolvable ambiguity whether the material is located closer or farther to us than the star. In other words, we assume that the dust is located in front of the ``plane of the sky'', defined as the imaginary plane perpendicular to the line of sight passing through the central star, rather than behind.

We also make the assumption that the stellar light is scattered only once on the dust grains before reaching us. This is clearly a simplification, but this hypothesis should be valid for the inner parts of the disk, that exhibit the strongest scattered light surface brightness. Since the scattering from the disk is generated on the inside of the bipolar cone, and we are on the outside of the cone, the emission from the inner surface could likely be rescattered in the passage through the cone walls. However, the dust column density in these walls is apparently small, and the effect of this second forward scattering of the already polarized light from the inner disk is therefore not expected to change significantly the degree of linear polarization.
Another simplification of our computation of the scattering angle $\theta$ is that we consider both stars A and B together as a single source. As we are mostly interested in scattering that occurs in the disk and other features at a typical radius of 10\,AU, the two stars are mostly seen as a single source by the light scattering dust. Moreover, the flux contribution of B is limited to less than a quarter of that of the AGB star in the ZIMPOL wavelength bands.

Knowing the projected radius $r$ (in AU) of each point in the disk relative to the central star (from the angular separation and the parallax) and the scattering angle $\theta$, we can compute its altitude expressed (in AU) in front of the plane of the sky:
\begin{equation}
z = r\,\tan \left( \frac{\pi}{2} - \theta \right).
\end{equation}
To compute $\theta$, we first average the $p_L$ maps obtained with and without coronagraph, that gave very similar polarization degrees (Fig.~\ref{deconv80}), and then computed the altitude maps separately for each camera. The values of $z$ have been computed only where the SNR of $p_L$ is larger than 5, and $p_L > 0.5\%$.

\begin{figure}[]
        \centering
        \includegraphics[width=7cm]{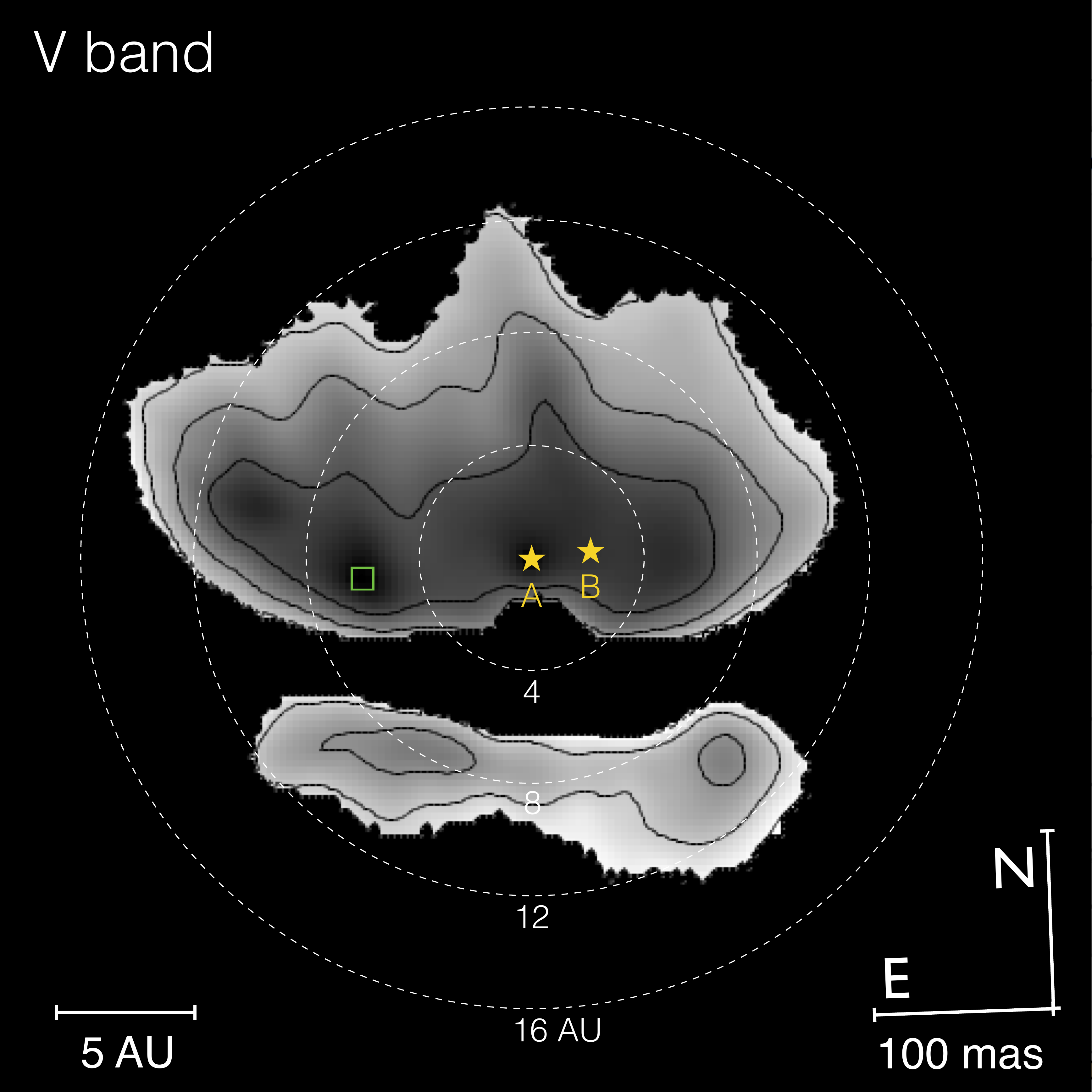}
        \includegraphics[width=7cm]{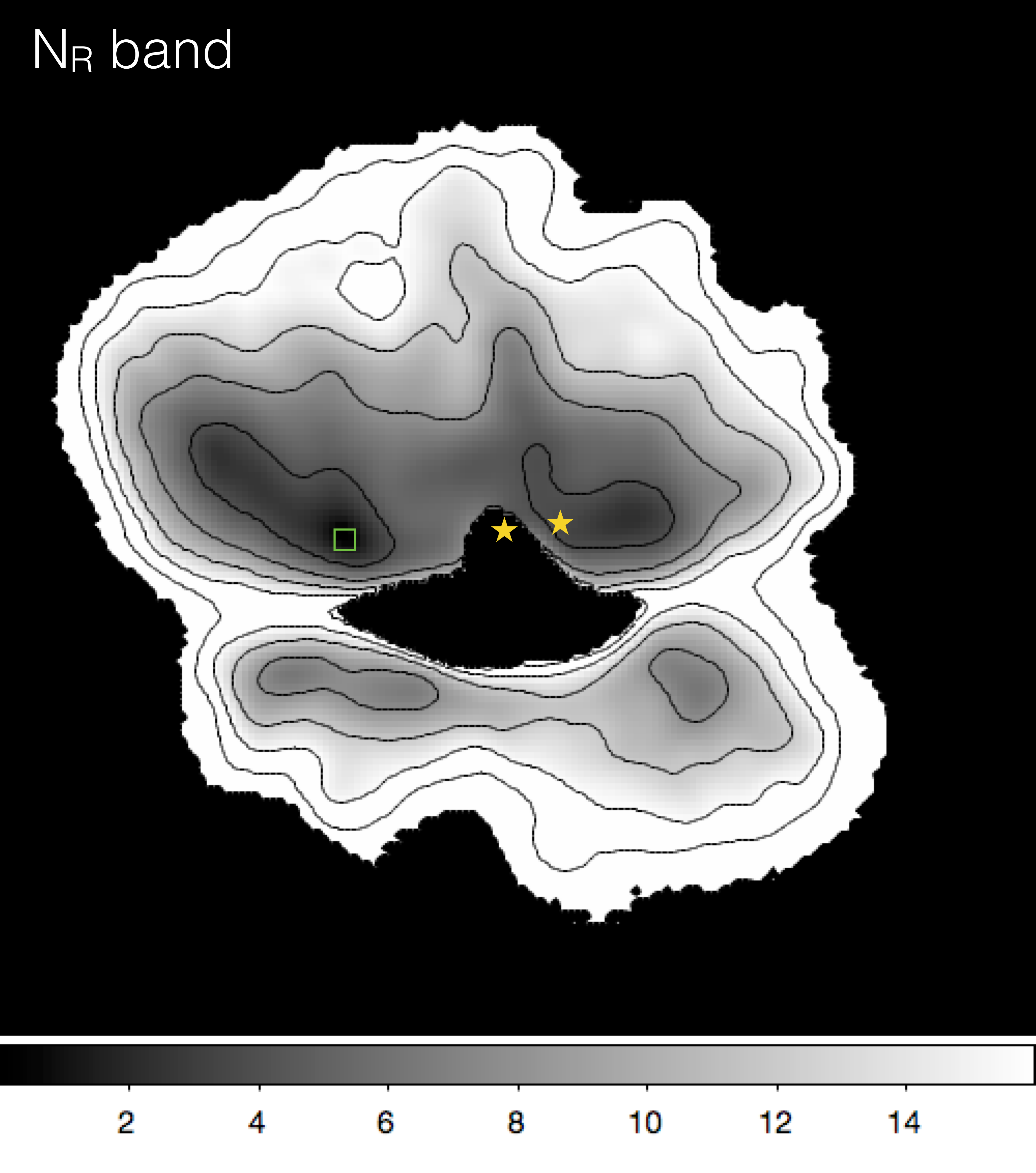}
        \caption{Map of the altitude of the scattering material in front of the plane of the sky at the distance of L$_2$\,Pup, expressed in astronomical units. The top panel shows the map derived from the $V$ band observations while the lower panel shows the $N_R$ band result. The positions of the AGB star and its companion are marked with star symbols, and the location of minimum altitude is shown with a green square. The gray scale is linear between 0 and 16\,AU, and the contour curves are separated by increments of 4\,AU from 0 to 20\,AU. The field of view is $0.60\arcsec \times 0.60\arcsec$.
        \label{zmap}}
\end{figure}

The results are shown in Fig.~\ref{zmap}. We observe a deep minimum of $z$ on the eastern side of the A star, that we interpret in Sect.~\ref{discussion} as the intersection of the inner rim of the dust disk with the plane of the sky. A symmetric minimum with respect to A is also present. There is a good match between the maps obtained for the two filters, although the map in the $N_R$ band extends slightly further than the map in the $V$ band, thanks to the higher brightness of the star and better polarizing efficiency of the dust in the $N_R$ band.

\subsection{RADMC-3D radiative transfer modeling \label{radmc}}

\subsubsection{Dust disk parameters}

To determine the physical parameters of the disk of L$_2$\,Pup, we adopt as a starting point the disk model presented by \citetads{2014A&A...564A..88K}, that is based on the RADMC-3D radiative transfer modeling tool \citepads{2012ascl.soft02015D}, in its version 0.38.
This code first computes the dust temperature using a thermal Monte Carlo simulation \citepads{2001ApJ...554..615B}, then it produces an image using isotropic scattering. The parameters of the regular grid we used are listed in Table~\ref{grid_radmc_table}. We did not use grid refinement.

\begin{table}
	\caption{Grid parameters of the RADMC-3D simulation}
	\centering          
	\label{grid_radmc_table}
	\begin{tabular}{ll}
		\hline\hline
		Parameter & Value \\
		\hline         
		\noalign{\smallskip}
		Inner radius & 3 AU \\
		Outer radius & 100 AU \\
		Number of points in radius direction & 500 \\
		Maximum meridional angle & 0.8 rad \\
		Number of points in meridional angle & 120 \\
		Azimuthal direction & Invariant \\	
		Wavelength interval limits ($\mu$m) & $\left[0.1,1,10,100 \right]$ \\
		Number of points in each interval & $\left[100,50,20 \right]$ \\
		\hline
	\end{tabular}
\end{table}

As the primary component L$_2$\,Pup~A dominates photometrically over B in both the $V$ and $N_R$ filters (Sect.~\ref{photometry-astrometry}), and to simplify the computation while preserving the axial symmetry, we considered the AGB primary as the single central source of the model. Consistently with \citep{2014A&A...564A..88K} we used a stellar radius of 123~R$_\odot$, and the spectral energy distribution (SED) was modeled using the Castelli-Kurucz library\footnote{Available at \url{http://www.stsci.edu/hst/observatory/crds/castelli_kurucz_atlas.html}} \citepads{2004astro.ph..5087C} with the following parameters: $T_\mathrm{eff}$ = 3500~K, $\log g$ = 0.0 and $\left[M/H\right]$ = 0.0.
We allowed for a vertical shift of the central star relative to the dust disk plane, and we note z$_\mathrm{\star}$ this offset, counted positively to the north.
We used a flared disk dust distribution described by the height radial distribution:
\begin{equation}
	h(r) = hr_\mathrm{out} \times r \times \left(\frac{r}{r_\mathrm{grid, \, out}}\right)^\beta
\end{equation}
Where $hr_\mathrm{out}$ is the ratio between $h$ and $r$ at the outer radius of the grid $r_\mathrm{grid, \, out}$, and $\beta$ the flaring index. The assumed dust density distribution is:
\begin{equation}
	\rho(r, \theta) = \rho_r (r)\  \rho_\theta (r, \theta)
\end{equation}
with:
\begin{equation}
	\rho_r (r) = \left[\sigma_\mathrm{out} \left(\frac{r}{r_\mathrm{out}}\right)^\alpha\right] / 
	\left[h(r)\sqrt{2\pi}\right],
\end{equation}
where $\sigma_\mathrm{out}$ is the dust density at the external radius of the disk $r_\mathrm{out}$ and:
\begin{equation}
	\rho_\theta (r, \theta) = \exp\left(-0.5 \left[ \frac{r}{h(r)} \tan\left( \frac{\pi}{2} - \theta \right) \right]^2 \right),
\end{equation}
$\theta$ being the meridional angle.
Using the model parameters derived by \citetads{2014A&A...564A..88K} as our initial guess, we reproduce morphologically and photometrically the observations in the ZIMPOL filters.
We used two amorphous dust species: MgFeSiO$_4$ (olivine) and Mg$_{0.5}$Fe$_{0.5}$SiO$_3$ (pyroxene) whose specific weights are respectively 3.7 and 3.2 g cm$^{-3}$. These species are among the most abundant in dust-forming circumstellar dust shells \citepads{2006A&A...447..553F}, and were already selected for the modeling by \citetads{2014A&A...564A..88K}.
Their physical properties\footnote{Available at \url{http://www.astro.uni-jena.de/Laboratory/OCDB/amsilicates.html}} were taken from \citetads{1994A&A...292..641J} and \citetads{1995A&A...300..503D}.

\begin{table}
	\caption{Derived parameters of the RADMC-3D dust disk model.}
	\label{result_RADMC_table}      
	\centering          
	\begin{tabular}{ll} 
		\hline\hline
		Derived parameter & Value \\
		\hline
		\noalign{\smallskip}
		z$_\mathrm{\star}$ & -0.5~AU\\
		R$_\mathrm{in}$ & 6~AU \\
		R$_\mathrm{out}$ & 13~AU \\
		$\alpha$ & -3.5 \\
		MgFeSiO$_4$ density at R$_\mathrm{out}$ & $5\ 10^{-4}$~g\,cm$^{-2}$ \\
		MgFeSiO$_4$ grain size & 0.1~$\mu$m \\
		MgFeSi$_2$O$_3$ density at R$_\mathrm{out}$ & $9\ 10^{-4}$~g\,cm$^{-2}$ \\
		MgFeSi$_2$O$_3$ grain size & 0.3~$\mu$m \\
		$hr_\mathrm{out}$ & 0.3 \\
		$\beta$ & 0.8\\
		Disk inclination & 82$^\circ$\\
		Total dust mass & $2.4\ 10^{-7}$~M$_\odot$ \\
		\hline
	\end{tabular}
\end{table}

\begin{figure}
        \centering
        \includegraphics[width=4.4cm]{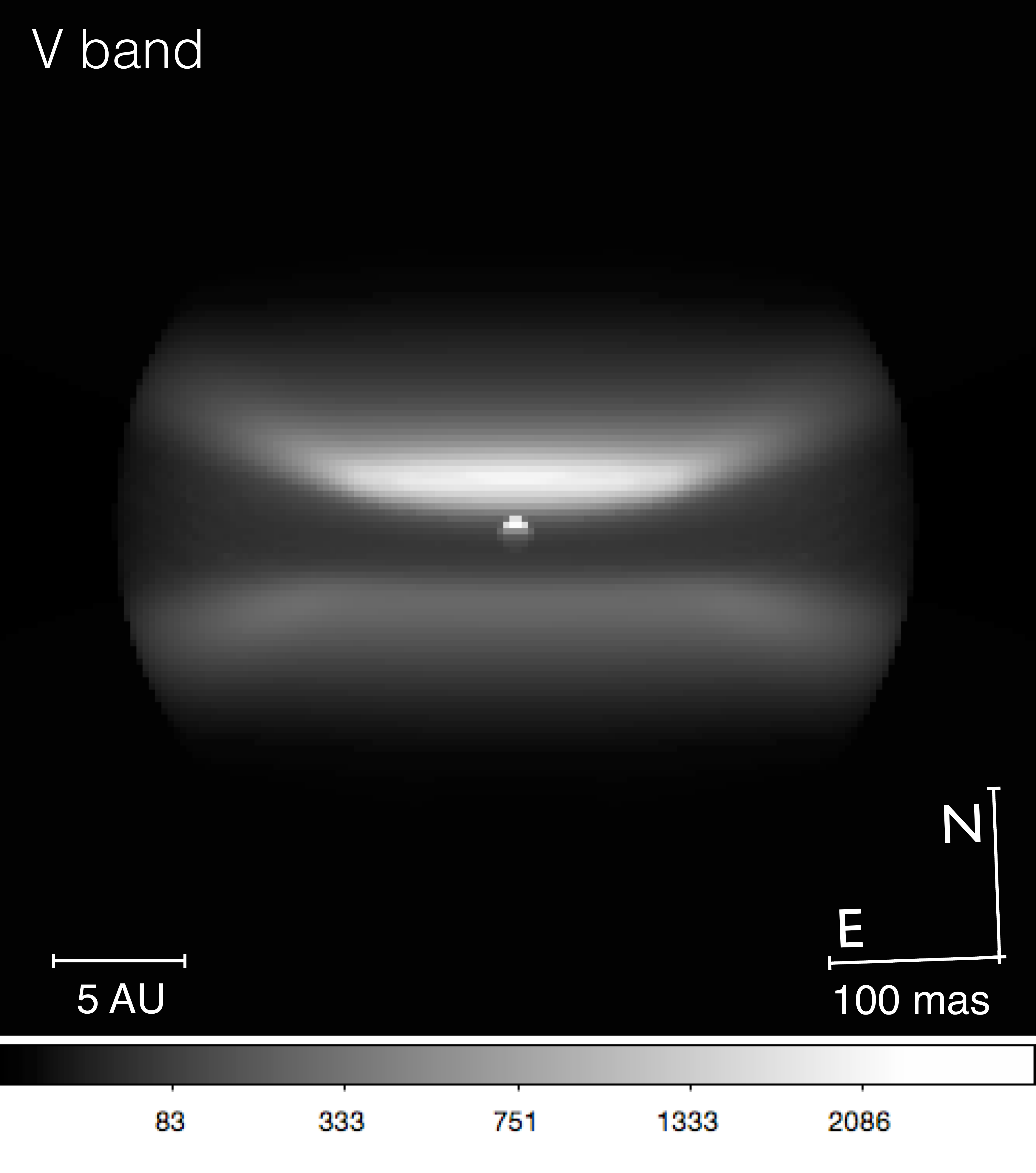}
        \includegraphics[width=4.4cm]{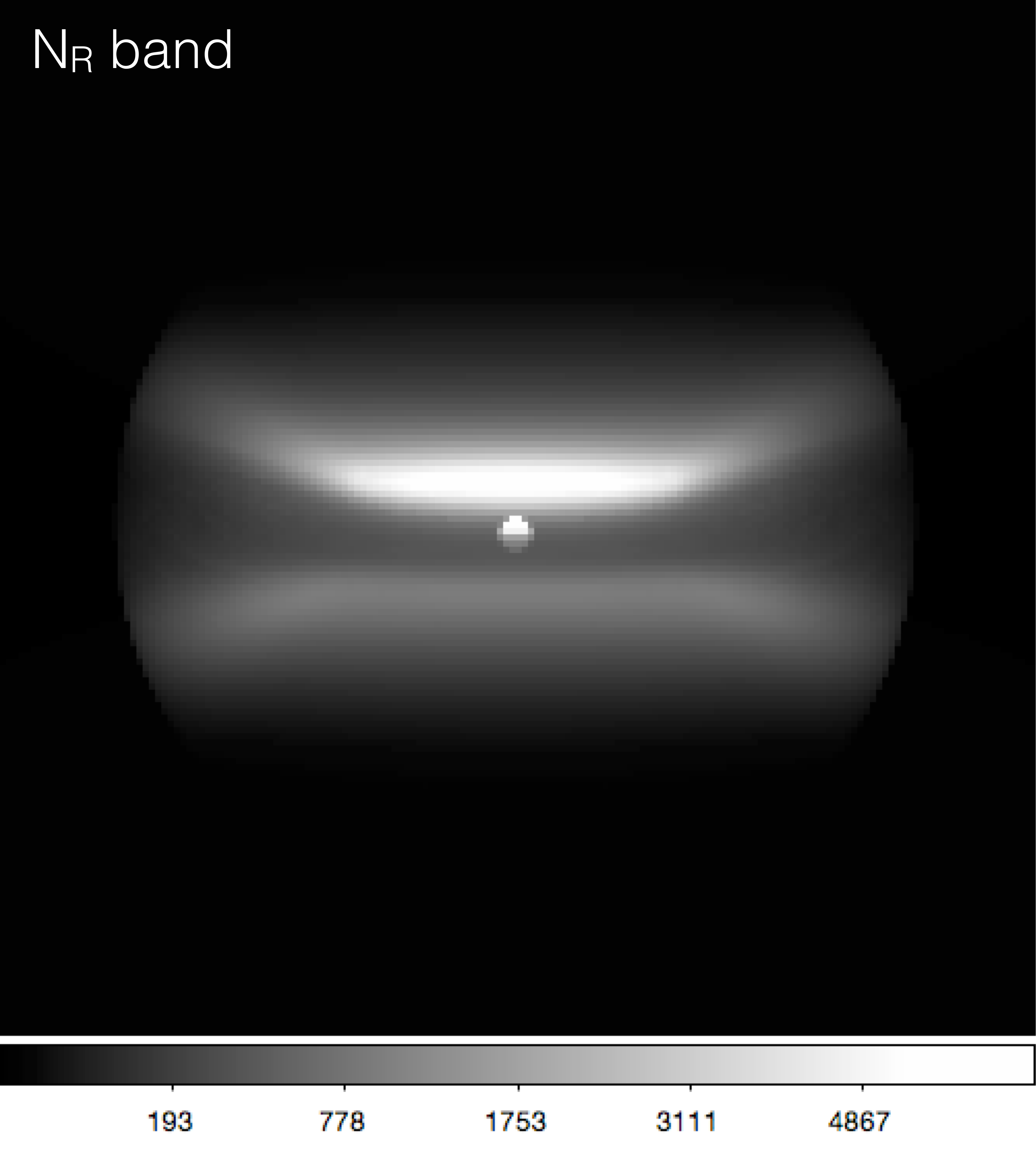}
        \caption{RADMC-3D Model image of the circumstellar disk of L$_2$\,Pup in the $V$ (left) and $N_R$ (right) bands.  The color scale is a function of the square root of the surface brightness (in Jy\ arcsec$^{-2}$).
        \label{radmc-model}}
\end{figure}

The model parameters giving the best agreement with the images and SED are listed in Table \ref{result_RADMC_table}, and the corresponding model images are represented on Fig.~\ref{radmc-model}. As the RADMC-3D model is a pure flared disk, it does not reproduce the asymmetric features discussed in Sect.~\ref{nebula-features} (spirals, plumes). We do not list confidence intervals for the parameters, as due to their relatively large number and the computing time required to generate a model, it is not possible to directly fit all the model parameters together to the observables (images and SED).
However, with the two selected dust species, the solution is well defined. Changing the grain size significantly degrades the fit. For instance, adding smaller grains results in the dust becoming too transparent in the red and infrared. The overall shape of the SED is also very sensitive to the inclination of the disk. The fluxes in the visible and in the infrared are well constrained by the dust densities and the visibility of the stellar apparent disk. To match the photometry, we need to partly see the stellar disk above the main dust band in the $N_R$ band, as well as the light scattered on the inner rim of the disk. But the star has to remain partially obscured for the flux to stay at a sufficiently low level, particularly in the near infrared.
The number of dust species remains however relatively uncertain. Considering the NACO images alone, only the two selected species could reproduce the flat shape of the SED from 1 to 4\,$\mu$m, but additional species may also be present.

The total dust mass in the RADMC-3D disk model is $2.4\ 10^{-7}$~M$_\odot$. This low mass (equivalent to $0.08$\,M$_\oplus$) however results in a relatively high dust density in the disk due to its small geometrical dimensions. Assuming a gas-to-dust ratio of the order of 100, this dust mass translates to a total mass including gas of $10^{-5}$ to $10^{-4}$~M$_\odot$. Considering the mass loss rate of $10^{-9}$ to $10^{-8}$~M$_\odot$\,yr$^{-1}$ estimated by \citetads{2002A&A...388..609W} from CO line observations in the millimeter domain, this corresponds to approximately 10\,000\,years of mass loss of L$_2$\,Pup at its current rate. As we have only a limited knowledge of the physical properties of L$_2$\,Pup\,B, we do not exclude that it may have contributed part of the dust present in the disk.

Compared to \citetads{2014A&A...564A..88K}, we obtain a slightly lower disk inclination (82$^\circ$ instead of 84$^\circ$), that may be partly due to the fact that we place the central star slightly below the disk plane (by approximately one stellar radius).
We also find a significantly larger disk thickness, as the efficient scattering in the visible and the high resolution of the ZIMPOL images allow us to see the dust band more clearly than in the infrared with NACO.
In the model by \citetads{2014A&A...564A..88K}, the density of the dust at a radius of 13\,AU is $\sigma_\mathrm{out} = 1.67\ 10^{-4}\,\mathrm{g\,cm}^{-2}$ for MgFeSiO$_4$ and $\sigma_\mathrm{out} =2.39\ 10^{-4}\,\mathrm{g\,cm}^{-2}$ for MgFeSi$_2$O$_3$. These densities are approximately 4 times smaller than those found in the present model at the same radius R$_\mathrm{out}$ (Table~\ref{result_RADMC_table}). This difference can be explained by the higher visible obscuration that has to be included in the model to match the ZIMPOL photometry.
The outer radius of the disk is smaller than in \citetads{2014A&A...564A..88K} at 13\,AU. However, we do not exclude that cold dust is present at larger distances, as the spectral energy distribution of L$_2$\,Pup shows excess up to the radio domain.

\subsubsection{Spectral energy distribution}

\begin{figure*}
        \centering
        \includegraphics[width=15cm]{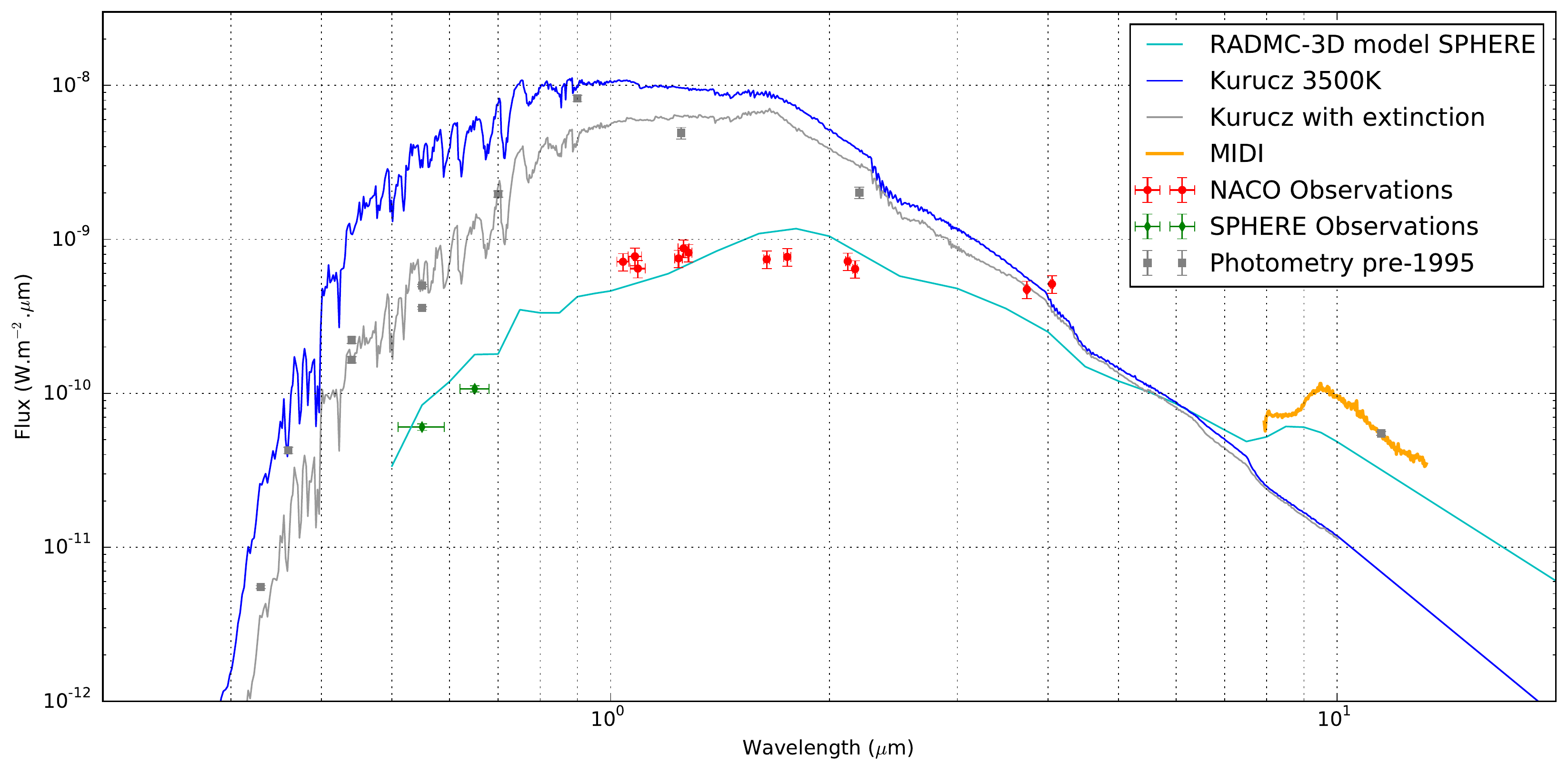}
        \caption{Spectral energy distribution of L$_2$\,Pup.
        \label{sed-model}}
\end{figure*}

The integrated spectral energy distribution (SED) of L$_2$\,Pup and its envelope is presented in Fig.~\ref{sed-model}.
Our new ZIMPOL photometry is shown using green points.
The early 2014 NACO photometry is represented by red points, while the orange curve is the MIDI spectrum \citepads{2014A&A...564A..88K}.
The gray points and curve represent the pre-dimming photometry (before 1995) and the reddened Kurucz model with $E(B-V)=0.6$, respectively.
The SED prediction by our RADMC-3D model is represented as a cyan curve, and the Kurucz model (L$_2$\,Pup~A only) used for our modeling as a blue curve.

Before the long-term fading that started in 1995, the SED was relatively well modeled from the ultraviolet to the near-infrared by a simple extinction of the Kurucz stellar SED ($T_\mathrm{eff}=3500$\,K) by Milky Way dust with $E(B-V)=0.6$, as demonstrated by the gray curve and gray points in Fig.~\ref{sed-model}. In this basic approach, the thermal emission from the disk was naturally not included.
For our recent (post-fading) photometry from NACO and ZIMPOL (obtained in early and late 2014), it is not possible to match both the near-infrared and visible photometry using such a simple model, as $E(B-V) \approx 1.5$ predicts correctly the visible photometry but is insufficient for the near-infrared.

The RADMC-3D model by \citetads{2014A&A...564A..88K} reproduced properly the flat shape of the SED of L$_2$\,Pup between 1 and 4\,$\mu$m, as well as the silicate bump around $\lambda=10\,\mu$m and the thermal emission tail at longer wavelengths. As shown in Fig.~\ref{sed-model}, our current model gives a predicted flux at visible wavelengths that is compatible with the ZIMPOL measurements. To reach this agreement, we adjusted the model parameters (disk thickness and inclination in particular), but the assumed dust properties are the same (composition, grain size).

\section{Discussion}\label{discussion}

\subsection{L$_2$\,Pup A and B\label{companion_properties}}

Considering the observed flux ratio between the fainter companion L$_2$\,Pup~B and the AGB star A in the visible ($\approx 24\%$), the companion is probably a red giant star, of somewhat smaller mass as A and in a less evolved state.
Without further spectral information on L$_2$\,Pup~B, we cannot exclude that it is of non-stellar nature (e.g. a prominent dust knot). However, the high relative flux compared to the AGB star favors the stellar companion hypothesis, that we adopt in the following.
The color of B, $(V-R)_B=1.54$, is slightly bluer than that of A, $(V-R)_A=1.79$, which is consistent with this interpretation. It is however difficult to assess precisely the spectral type of the companion based on its color alone, due to the strong circumstellar reddening.

Another approach to determine the properties of the secondary is to consider the color excess $E(B-V)=0.6$ determined by \citetads{2014A&A...564A..88K} based on the photometry obtained before the fading of 1995. Using this parameter, we correct the apparent $(V-R)$ colors of A and B (Table~\ref{photom_table}), and we obtain $(V-R)_A=1.2$ and $(V-R)_B=1.0$. These colors correspond to an early M giant for A (consistent with its effective temperature) and a late K (or very early M) giant for B. The mass of L$_2$\,Pup~B could be approximately of $1.5\,M_\odot$, as considering the rapidity of giant branch (GB) evolution, to have a stellar pair on the GB and AGB simultaneously implies that they were close in mass while on the main sequence.

To determine the plausible orbital period of the central binary, we consider the classical relation from Kepler's third law:
\begin{equation}
P_\mathrm{orb}^2 = \frac{4 \pi^2 r^3}{G\left( m_1 + m_2 \right)},\end{equation}
where $m_1$ and $m_2$ are the two object masses, $P_\mathrm{orb}$ the orbital period, and $r$ the separation between the stars (the orbits are assumed to be circular). For equal mass components of 1.5 to 2\,M$_\odot$, and a separation of 2 to 4\,AU, we obtain orbital periods of $P_\mathrm{orb} = 1.4$ to 4.6\,years. To obtain a more precise estimate, we plan to follow the apparent stellar positions through part or all of the binary period.
We note that the mass of L$_2$\,Pup is currently uncertain, as \citetads{lykou15} derive a low value of $0.7\,M_\odot$ for the AGB star from the dynamics of the SiO masers \citepads{2013ApJ...774...21M}, while \citetads{1998NewA....3..137D} and \citetads{2014A&A...564A..88K} propose values closer to 2\,$M_\odot$. The influence of the presence of a second star in the system on these estimates is unclear. We postpone a more detailed discussion of the evolutionary state of the L$_2$\,Pup system to a forthcoming paper.

\subsection{Overall disk geometry}

\begin{figure*}
        \centering
        \frame{\includegraphics[width=4.3cm]{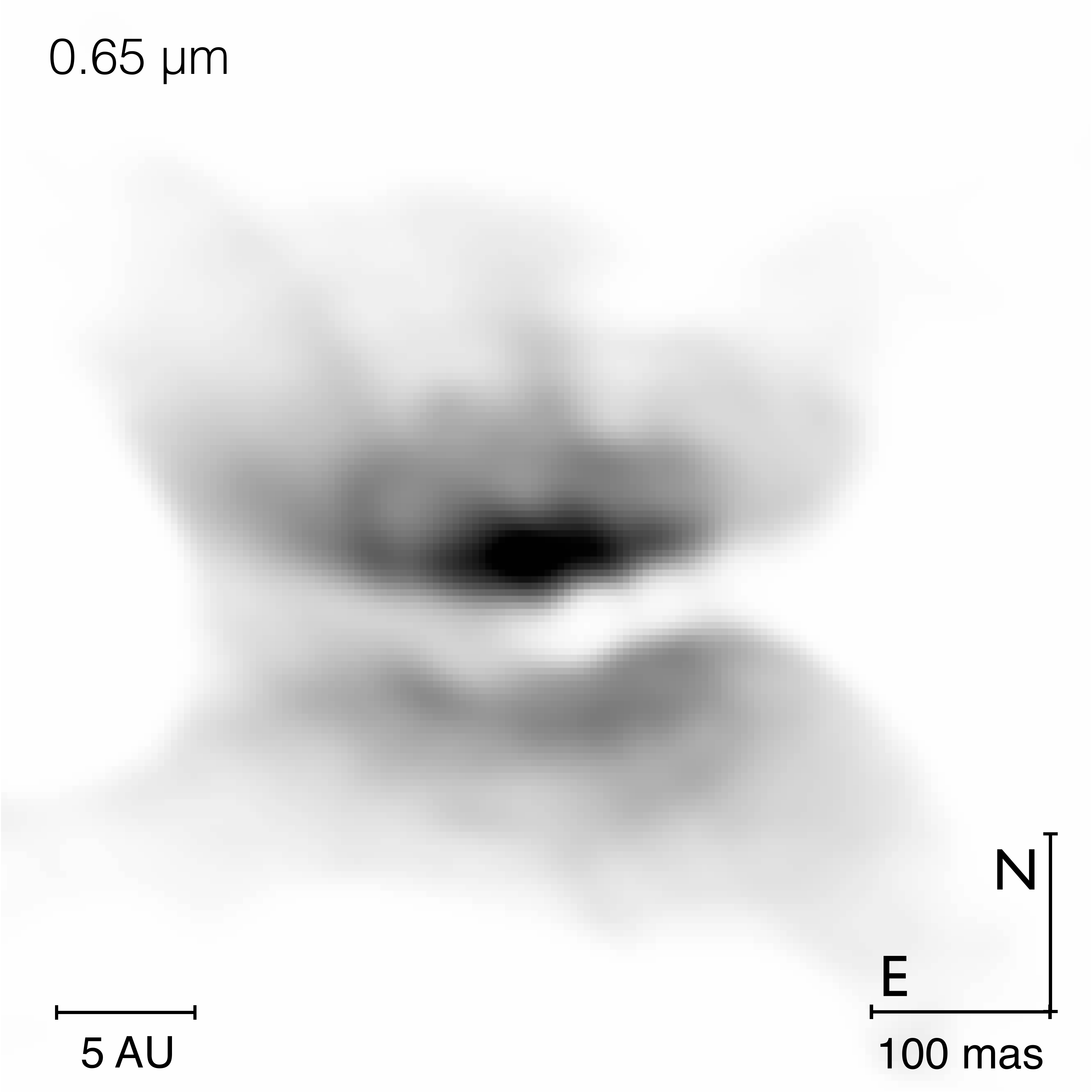}}
        \frame{\includegraphics[width=4.3cm]{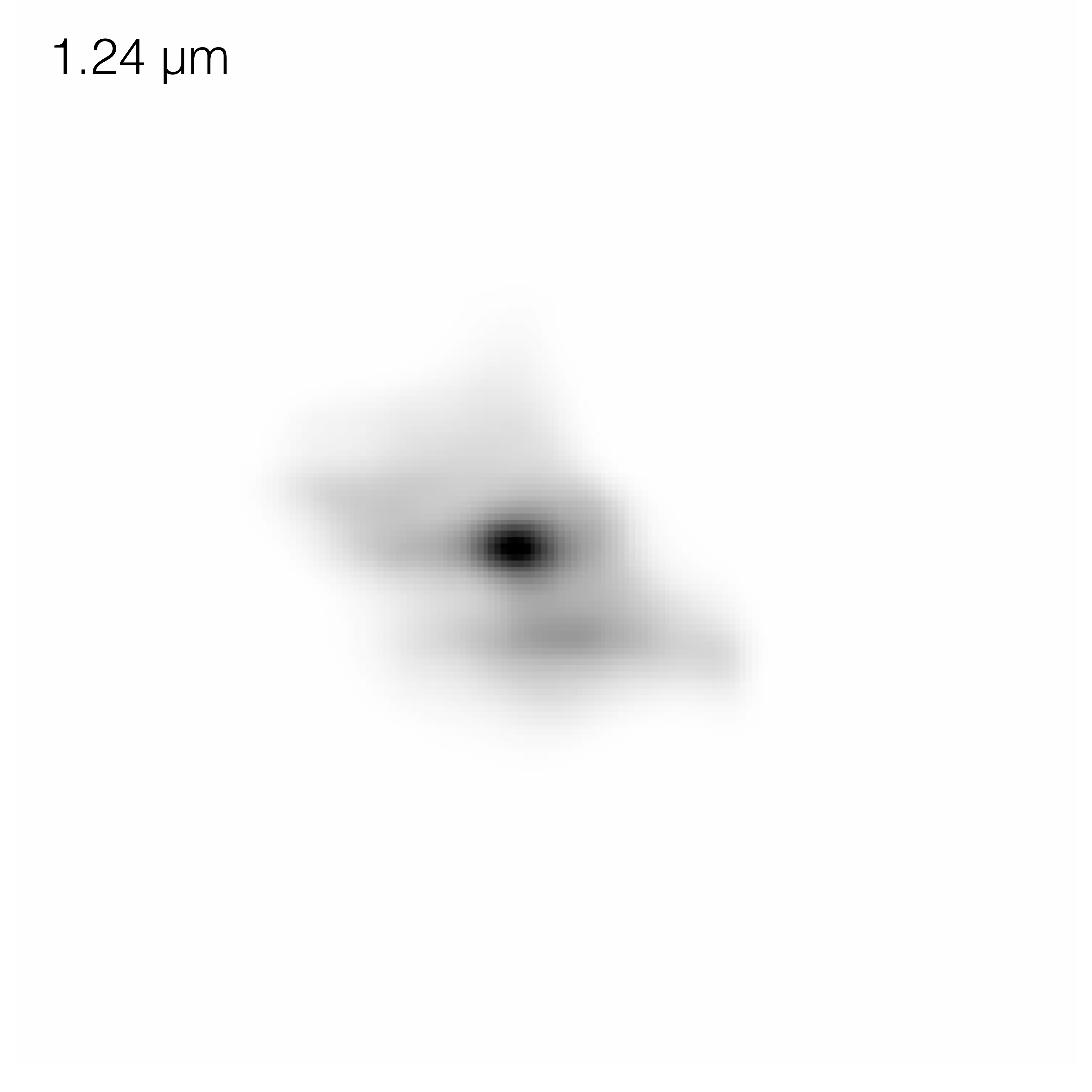}}
        \frame{\includegraphics[width=4.3cm]{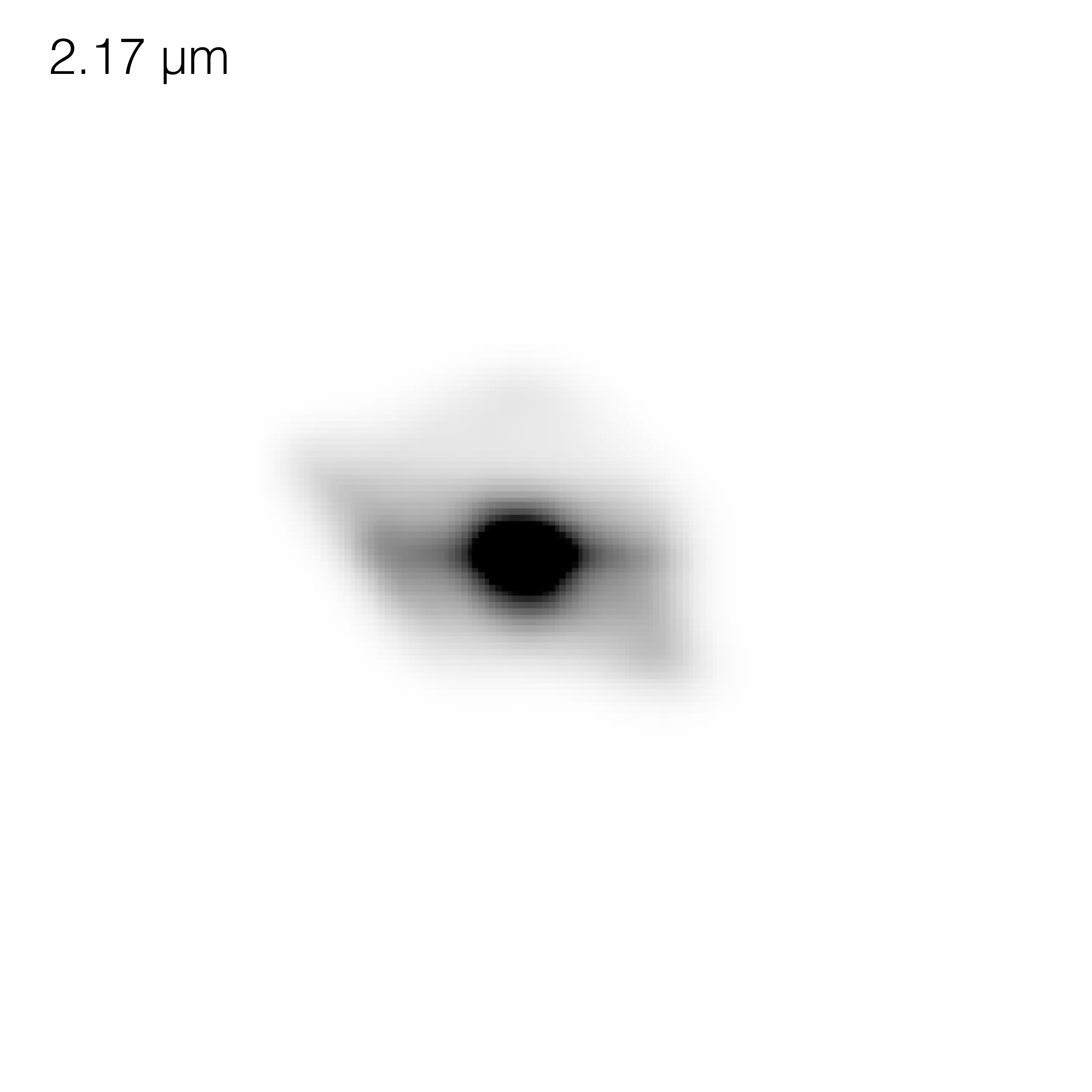}}
        \frame{\includegraphics[width=4.3cm]{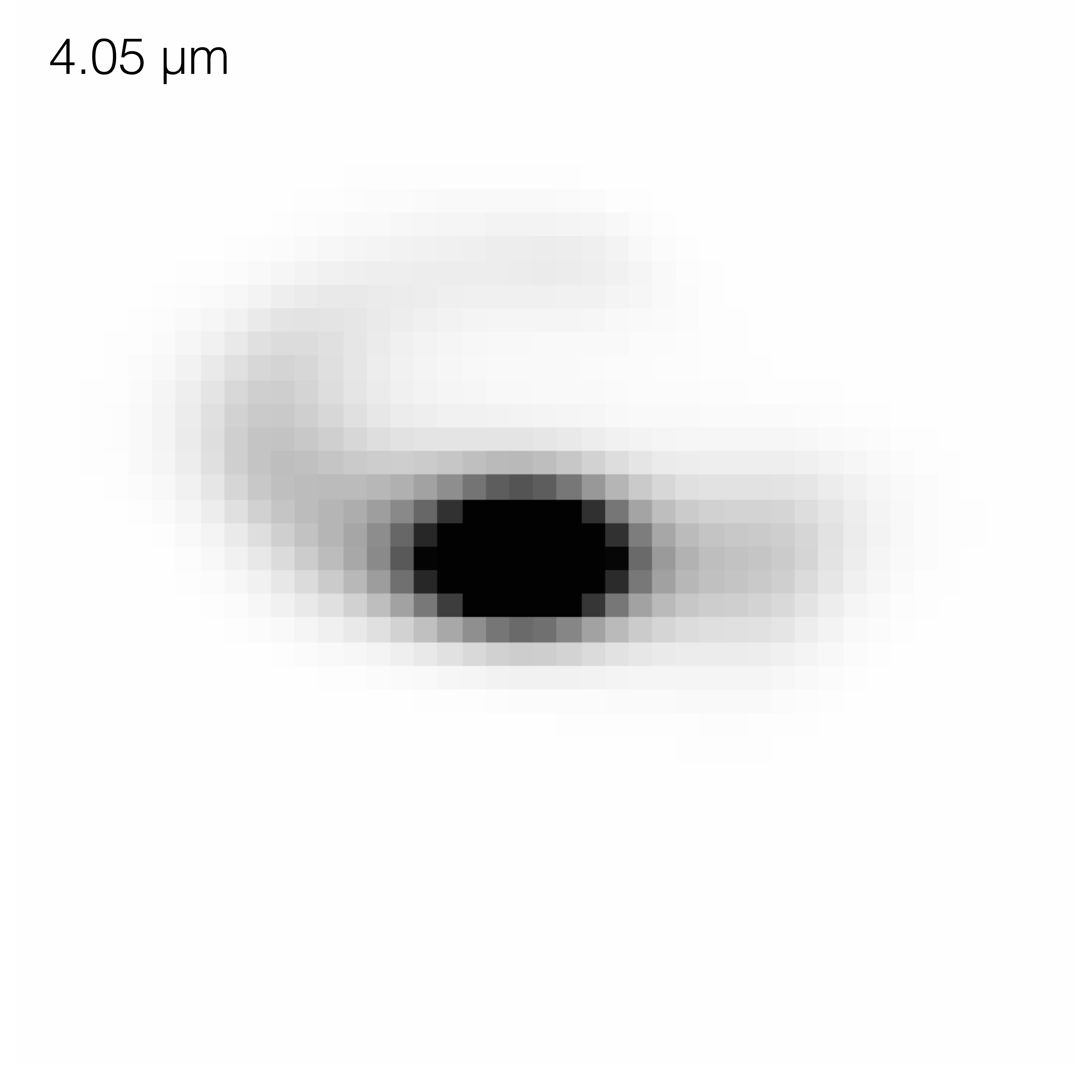}}
        \caption{Visible (ZIMPOL, logarithmic intensity scale) and infrared \citepads[NACO, from][]{2014A&A...564A..88K} images (square root scale) of L$_2$\,Pup.
        \label{visible-infrared}}
\end{figure*}

A comparison of the $N_R$ light intensity map with the infrared images from \citetads{2014A&A...564A..88K} is presented in Fig.~\ref{visible-infrared}.
We interpret the compact polarization maximum visible to the east of L$_2$\,Pup~A (marked with a green square in Fig.~\ref{zmap}), as the inner rim of the disk. It is consistently detected in both the $V$ and $N_R$ band polarimetric maps, and it is interesting to remark that another $p_L$ maximum is present almost symmetrically with respect to L$_2$\,Pup~A. These polarization maxima and the central star are aligned with the major axis of the disk.
The radius of the $p_L$ minimum measured from L$_2$\,Pup~A is R$_\mathrm{rim} = 6.0 \pm 0.2$\,AU. This value is in excellent agreement with the inner rim radius of 6\,AU proposed by \citetads{2014A&A...564A..88K} based on the $K$ band image: the thermal emission from the inner rim of the disk can be seen as an east-west segment in the $2.17\,\mu$m image presented in Fig.~\ref{visible-infrared}. This radius is also consistent with the corresponding RADMC-3D model parameter presented by these authors to reproduce the SED. 
The semi major axis of the ellipse represented in Fig.~\ref{features} is R$_\mathrm{ext} = 13$\,AU, and the thickness of the dark dust band is $h_\mathrm{ext} = 3$\,AU. We propose that $R_\mathrm{ext}$ corresponds to the external radius of the dense part of the dust disk and $h_\mathrm{ext}$ is its thickness at its external radius.

It is interesting to remark that the values of R$_\mathrm{rim}$, R$_\mathrm{ext}$ and $h_\mathrm{ext}$ that were directly measured from the ZIMPOL intensity and $p_L$ maps are respectively in very good agreement with the best fit parameters R$_\mathrm{in}=6$\,AU, $R_\mathrm{out} = 13$\,AU and $hr_\mathrm{out} \times R_\mathrm{out} = 4$\,AU listed in Table~\ref{result_RADMC_table}. With the selected dust mixture and grain size, they are the only combination that produces model images and SED matching the observed ones.
This convergence gives confidence that the RADMC-3D model geometry is a reasonably good match to the true dust distribution around L$_2$\,Pup.

We can check the compatibility of the radius of the inner rim of the disk with the presence of dust by computing the equilibrium temperature at this distance from the AGB star. We consider the absolute magnitude of L$_2$\,Pup before the fading that occurred after 1995 (see also Sect.~\ref{fading}), $M_K=-2.34$ and color $(V-K)=7.0$ \citepads{2002MNRAS.337...79B}.
We consider the $K$ -band angular diameter measurements of \citetads{2005ApJ...620..961M}, by using six stars with a similar $(V-K)$ color at phase 0.  The apparent diameter for $K=0$ is $7.5 \pm 1$\,mas, and the angular diameter of L$_2$\,Pup~A is thus estimated to $\approx 22 \pm 3$\,mas, in agreement with the $19 \pm 1.6$\,mas measurement by \citetads{2014A&A...564A..88K}.  As the radius of the inner edge of the disk is $94 \pm 4$\,mas, then the ratio of disk inner rim to stellar radius is $8.5 \pm 2$. Assuming the conservation of energy in a spherical geometry, the rim of the disk will have an equilibrium temperature $\sqrt{8.5}$ smaller than the effective temperature of the star.  The pre-dimming effective temperature of star A from \citetads{2002MNRAS.337...79B} is T$_\mathrm{eff} = 2900$\,K (we neglect the contribution of B), that gives $\mathrm{T} \approx 1000$\,K at the disk inner boundary.
Assuming a higher T$_\mathrm{eff} =3500$\,K for the star results in $\mathrm{T} \approx 1200$\,K at the inner rim.
In the absence of mineralogical information, these is a plausible range of temperature for dust formation \citepads[see e.g.][]{2013A&A...555A.119G}. 

We do not identify clear signatures of shadowing effects in the cones or on the disk from the inner system components \citepads[see e.g.][]{2015ApJ...798L..44M}, but such effects are difficult to disentangle from dust density fluctuations in a complex envelope such as that of L$_2$\,Pup. The overall asymmetry between the eastern and western parts of the cones may however be caused by differential shadowing effects, in particular if the inner rim of the disk has an inhomogeneous dust density and scale height.

\subsection{Specific features in the nebula \label{nebula-features}}

\begin{figure}
        \centering
        \frame{\includegraphics[width=8.5cm]{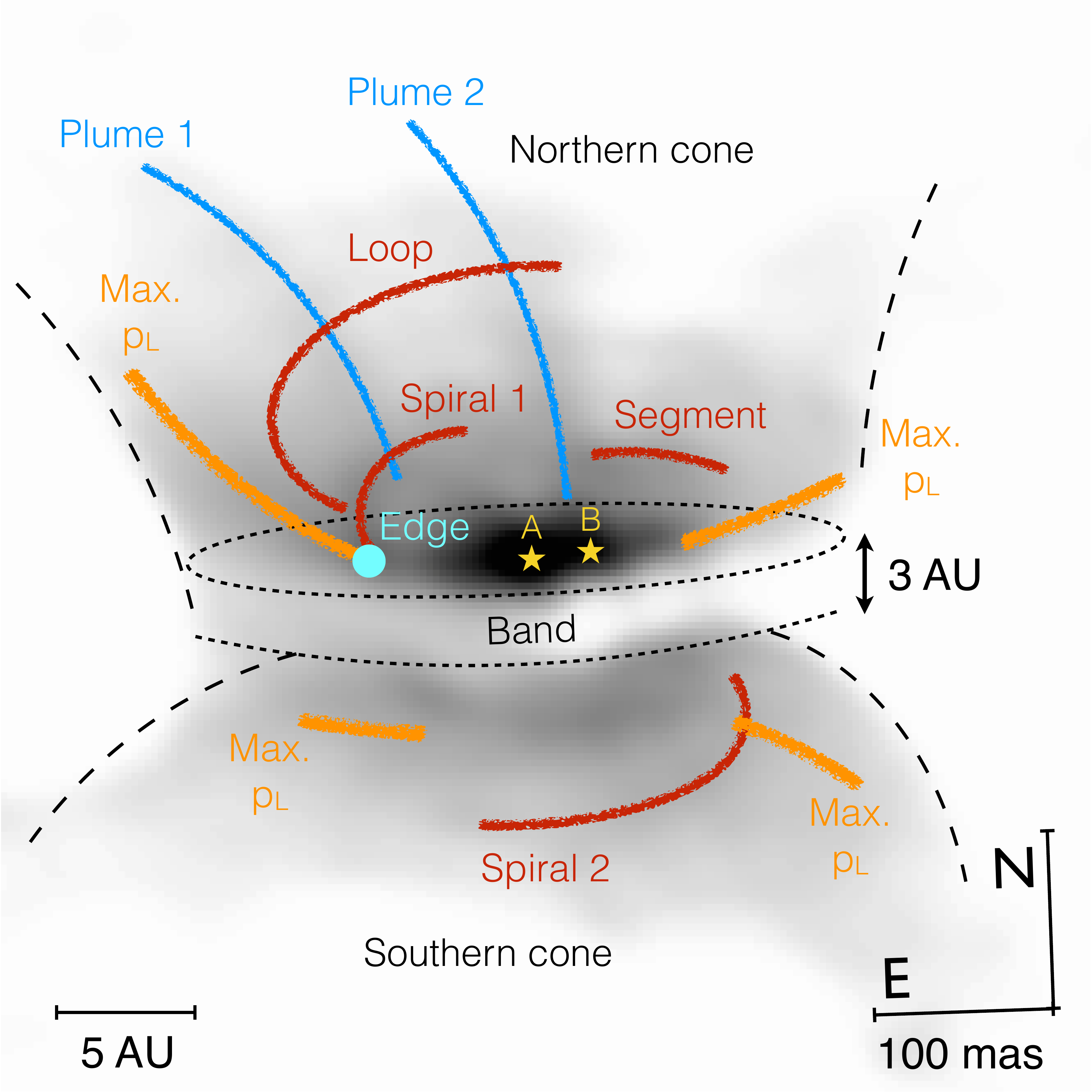}}
        \caption{Nomenclature of the features observed around L$_2$\,Pup.
        \label{features}}
\end{figure}

Several remarkable features are visible in the intensity maps obtained with ZIMPOL and NACO. A nomenclature of these features is presented in Fig.~\ref{features}. 

\subsubsection{The loop and spirals}

\begin{figure}
        \centering
        \frame{\includegraphics[width=8.5cm]{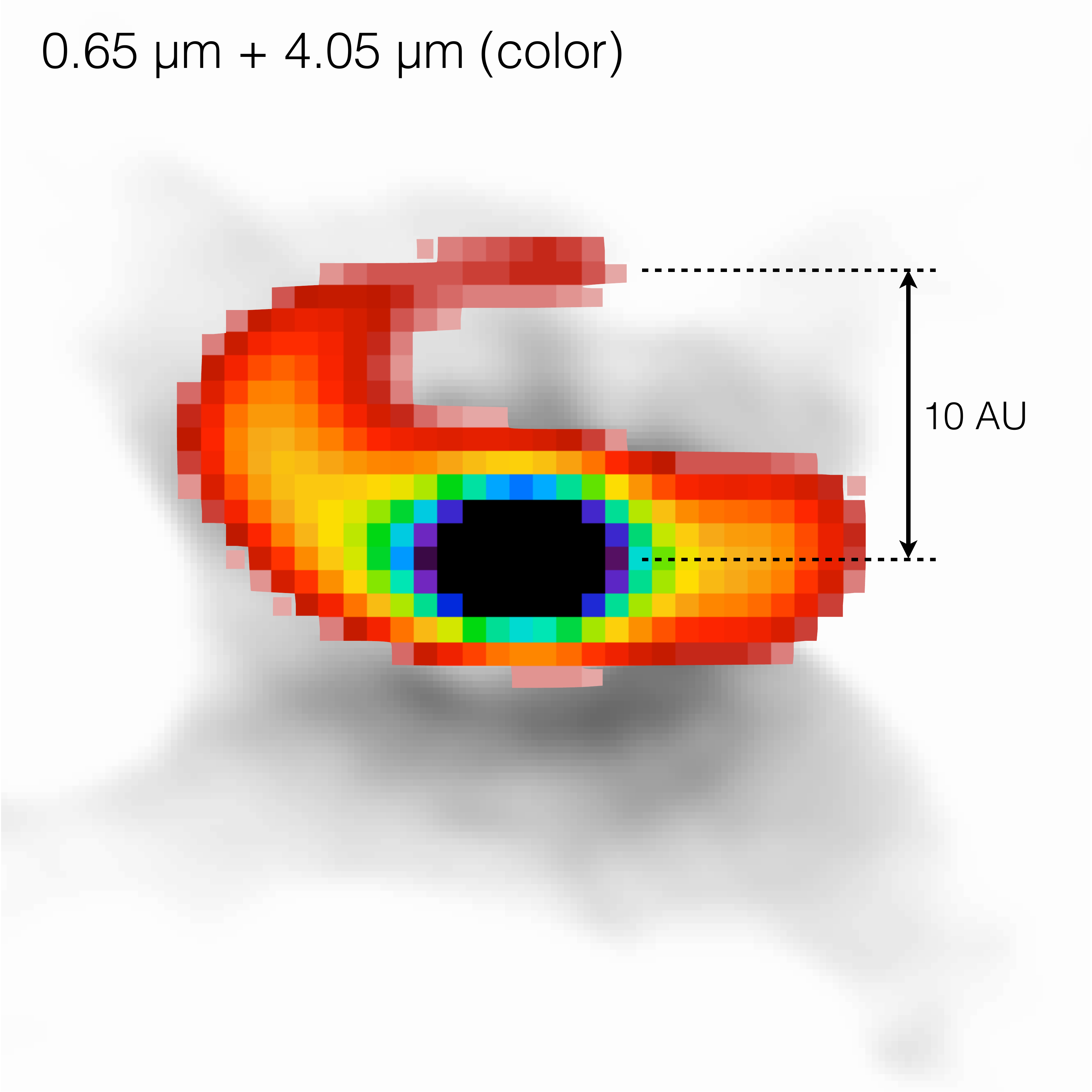}}
        \caption{$L$ band loop image from NACO (color scale) superimposed on the $V$ band ZIMPOL unpolarized intensity map (background black and white image). The field of view is $0.60\arcsec \times 0.60\arcsec$.
        \label{loop}}
\end{figure}

Following \citetads{2014A&A...564A..88K}, we define the \emph{loop} as the large thermal emission feature detected in the NACO $L$ band images. The loop is not clearly detected in the ZIMPOL unpolarized intensity maps. Its eastern part corresponds well to the maximum of the $p_L$ map presented in Fig.~\ref{deconv80}, and it appears to develop on the far side of the northern cone. Its non detection in visible scattered light is likely due to the inefficiency of backward scattering. A superposed view of the $N_R$ and thermal infrared $L$ band image at $4.05\,\mu$m is presented in Fig.~\ref{loop}, to show the relative positions of the loop and the northern cone.

A bright curved feature labeled spiral 1 in Fig.~\ref{features} is detected in the eastern part of the disk. Its apparent origin is located at the inner rim of the disk. Its true starting point may however be hidden behind the northern edge of the dust band. It is the brightest feature in the nebula apart from the two stars and the inner rim of the dust disk, and it is easily visible in the non-deconvolved images of L$_2$\,Pup (Fig.~\ref{nondeconv}).
We also note the presence of a second spiral (spiral 2 in Fig.~\ref{features}) in the southern cone of the nebula, whose winding orientation is the same as the loop and spiral 1, and whose position is mostly symmetric of the loop with respect to star A. Based on the $z$ map (Fig.~\ref{zmap}), the loop and also probably the two spirals seem to be physically located on the surface of the disk and cones.

The presence of L$_2$\,Pup's spirals indicates a behavior potentially similar to that observed in the binary AGB star R\,Scl by \citetads{2012Natur.490..232M} using ALMA. In both cases, spirals are observed, but the overall geometry of the envelope of R\,Scl is spherical, while the geometry of that of L$_2$\,Pup is essentially a bipolar structure.
We propose that the different geometry of L$_2$\,Pup is due to the presence of the dust disk that collimates the stellar wind along a specific spatial axis. \citetads{2012Natur.490..232M} deduced the recent mass loss history of R\,Scl from the spacing of the successive windings. Other examples exist, e.g.~\citetads{2014A&A...570L..14R} reported two-dimensional spirals in the Mira binary system. \citetads{2015A&A...575A..91C} identified a large spiral structure in the envelope of CW\,Leo at arcminute scales, using the 30-m IRAM antenna at millimeter wavelengths. \citetads{2015A&A...574A...5D} also detected arcs and a smaller spiral in the inner envelope of this star from ALMA, as well as a probable companion star from modeling of the spiral. 

In the case of L$_2$\,Pup, the spirals are apparently located on the internal surfaces of the cones, so this complicates the determination of the pitch of the spiral. The geometry of the $L$ band loop suggests that the initially two-dimensional spirals has been blown away by the stellar wind, collimated by the disk, to form two 3D streamers (to the north and south). We can roughly estimate the winding time scale of the loop from the spacing between the tip of the loop and the star. We measure a vertical spacing of 10\,AU between the central object and the northern end of the loop (Fig.~\ref{loop}).
The CO outflow velocity determined by \citetads{2002A&A...388..609W} is only 3 km/s, but with a collimated flow nearly orthogonal to the line of sight, we have to account for a projection effect. Considering the inclination of the disk ($\approx 82^\circ$, see Sect.~\ref{radmc}), we estimate the velocity of the outflow to $v_\mathrm{wind} = 3 / \cos(82^\circ) \approx 20$\,km/s. This value is faster than the 10\,km/s typical velocity of the material close to the photosphere derived by \citetads{2002A&A...388..609W} from the SiO masers. It is qualitatively in agreement with the wind velocity derived by \citetads{2012Natur.490..232M} for R\,Scl.
Considering $v_\mathrm{wind} = 20$\,km/s, the material in the loop takes $10\,\mathrm{AU} / (20\,\mathrm{km/s}) \approx 2.5$\,years to move vertically by one spiral winding. There is a large uncertainty on whether we observe a complete winding or only a partial one, but this duration is reasonably comparable with the plausible orbital period of the central binary (Sect.~\ref{companion_properties}).

\subsubsection{Plumes}

The extended plume~1 that emerges to the north of the disk has its apparent origin close to the maximum of polarization, i.e.~at the location that we identify as the inner rim of the dust disk, at a radius of 6\,AU. Its signature is detected in the polarization map, and its 3D orientation appears essentially perpendicular to the dust disk plane (Fig.~\ref{zmap}). The emission observed to the east of L$_2$\,Pup~A in Fig.~\ref{companion_photometry} could be the result of the wind and radiation pressure of B blowing the dusty wind of A toward the east. The similarity in position between the tip of spiral 1 and the start of plume 1 indicates that the plume may be ejected from this impact point, due to the local increased heating around the location of the shock.
Finally, another speculative explanation is that a low mass companion (possibly recently formed in the disk) is accreting material expelled from the AGB star, and ejects material through polar jets \citepads{2013MNRAS.435.2416V, 2012ApJ...744..136K}.

Mostly parallel to plume 1, plume 2 has also a clear signature in the $p_L$ maps (Fig.~\ref{deconv80}), and its 3D position appears again nearly perpendicular to the disk plane, as shown in Fig.~\ref{zmap}. Its origin can be traced in the inner binary system, and more precisely close to the position of the B star. One hypothesis is that plume 2 could originate from the collision of the winds of the two stars.  Another possibility is that part of the material expelled by the AGB component A is accreted through Roche lobe overflow on star B \citepads{2013A&A...552A..26A}, inducing the formation of an accretion disk. This process could then result in the appearance of polar enhanced mass loss streams. The faint X-ray emission detected by \citetads{2012A&A...543A.147R} would favor this scenario, although the authors indicate that it may be caused by the leak of red photons in the detector. 

\subsection{Long-term photometric variations of L$_2$\,Pup\label{fading}}

The visible flux of L$_2$\,Pup is subject to relatively large long-term variations of $\approx 2$ magnitudes in $V$ \citepads{2002MNRAS.337...79B}, with in particular a 5-year decrease in brightness that started around 1995, followed by a long and stable minimum. This kind of trend in Mira stars is rare, but not unknown \citepads{1991AJ....102..200D}. The time constant of such a trend seems too long to be associated with AGB pulsation, and to short to be related to AGB evolution.
Since the stars are obscured by the disk that is seen almost edge-on, we propose that the long-term photometric variations of L$_2$\,Pup are caused by changes in the line-of-sight opacity of the intervening material.
\citetads{2002MNRAS.337...79B} observed that the visible-infrared $(V-K)$ color increased significantly during the dimming event that started around the year 1995, which is consistent with the occultation (and associated reddening) by the dust present in the disk.
We also note that the ZIMPOL intensity images show that the edges of the main dust band exhibit an irregular shape, and that the shape of the band itself appears slightly curved in a ``V'' feature. The keplerian rotation of the disk therefore likely results in time variable occultation of the central stars.

\subsection{A future bipolar planetary nebula?}

The origin of bipolar planetary nebulae (PNe) is one of the great classic problems of modern astrophysics.  Bipolar PNe present a remarkable axial symmetry, whose origin is often attributed to the presence of a secondary star \citepads{2014arXiv1410.3692L, 1989ApJ...339..268S}, but a companion has been detectable only in few cases \citepads[e.g.][]{2015A&A...576L..15O}.
From its position in the HR diagram, L$_2$\,Pup appears as a relatively young AGB star, and therefore a possible progenitor of these spectacular structures.
Our ZIMPOL observations demonstrate that the close circumstellar dust is distributed in a dense disk, that also hosts a companion star in a tight orbit.

A possible explanation for the formation of the circumbinary disk is that part of the dust formed in the AGB star wind is confined in the orbital plane through interaction with the wind and radiation pressure of the secondary star.
Another fraction of the mass loss could then be collimated by the flared disk and accelerated orthogonally to the disk plane into a bipolar flow.
In this scenario, the presence of a companion and a collimating dust disk appear intimately tied together to explain the appearance of a privileged spatial axis in the mass loss.
An alternate hypothesis is that the companion accretes material ejected by the AGB star, forming an accretion disk and its associated jets. These jets would then carve a bipolar cavity in the slower AGB wind \citepads[see e.g.][]{2004ApJ...600..992G, 2006MNRAS.370.2004N}.
Based on the morphology of the envelope of L$_2$\,Pup that includes jets, streamers and a flared disk, both mechanisms may also be simultaneously active.
In this framework, it appears plausible that the L$_2$\,Pup system will evolve into a PN of the hourglass type.

\section{Conclusion}

The unprecedented angular resolution of the ZIMPOL images at visible wavelengths allowed us to confirm the discovery by \citetads{2014A&A...564A..88K} of a large dust disk around L$_2$\,Pup, and revealed the presence of a companion at a separation of only 33\,mas. 
The potential of the ZIMPOL instrument for the study of the close environment of objects enshrouded in dust appears excellent, thanks to the high efficiency of light scattering at visible wavelengths. Coupled with highly precise polarimetry, the three-dimensional structure of the envelopes can be determined with a high degree of fidelity through radiative transfer modeling.
Only very few binary AGB star disks have been identified to date. The disk we observed with ZIMPOL is therefore an important fiducial to improve our understanding of the formation of bipolar PNe.
The L$_2$\,Pup system is very favorable for observation, thanks to its proximity, and a companion is detected which is a significant indication for the generality of the binary case.

Following e.g.~the work of \citetads{2012Natur.490..232M} on R\,Scl, the observation of this system with ALMA will provide decisive information on the geometry and dynamics at play in this complex envelope. For instance, CO line observations will directly test our dust disk model through the monitoring of the keplerian motion of the gas in the disk. \citetads{2013ApJ...774...21M} identified a large velocity range in the SiO maser emission, as well as an intriguing three-peaked spectrum \citepads{2002A&A...388..609W, 2004A&A...422..651S}. Interferometric imaging with the exquisite angular resolution of the longest baselines of ALMA (comparable to NACO and ZIMPOL) will probe the near circumstellar region from where the SiO emission originates.

\begin{acknowledgements}
We are grateful to the SPHERE instrument team for the successful execution of our observations during the Science Verification of the instrument. We thank Dr.~Julien Milli for his help with the ZIMPOL data reduction pipeline.
We acknowledge financial support from the ``Programme National de Physique Stellaire" (PNPS) of CNRS/INSU, France.
STR acknowledges partial support by NASA grant NNH09AK731.
AG acknowledges support from FONDECYT grant 3130361.
This research received the support of PHASE, the high angular resolution partnership between ONERA, Observatoire de Paris, CNRS and University Denis Diderot Paris 7.
PK and AG acknowledge support of the French-Chilean exchange program ECOS-Sud/CONICYT.
This research made use of Astropy\footnote{Available at \url{http://www.astropy.org/}}, a community-developed core Python package for Astronomy \citepads{2013A&A...558A..33A}.
We used the SIMBAD and VIZIER databases at the CDS, Strasbourg (France), and NASA's Astrophysics Data System Bibliographic Services.
We used the IRAF package, distributed by the NOAO, which are operated by the Association of Universities for Research in Astronomy, Inc., under cooperative agreement with the National Science Foundation.
\end{acknowledgements}

\bibliographystyle{aa} 
\bibliography{biblioL2Pup}

\end{document}